# PREDICTIVE LEARNING VIA RULE ENSEMBLES

By Jerome H. Friedman [1] and Bogdan E. Popescu[2]

*Stanford University*

General regression and classification models are constructed as linear combinations of simple rules derived from the data. Each rule consists of a conjunction of a small number of simple statements concerning the values of individual input variables. These rule ensembles are shown to produce predictive accuracy comparable to the best methods. However, their principal advantage lies in interpretation. Because of its simple form, each rule is easy to understand, as is its influence on individual predictions, selected subsets of predictions, or globally over the entire space of joint input variable values. Similarly, the degree of relevance of the respective input variables can be assessed globally, locally in different regions of the input space, or at individual prediction points. Techniques are presented for automatically identifying those variables that are involved in interactions with other variables, the strength and degree of those interactions, as well as the identities of the other variables with which they interact. Graphical representations are used to visualize both main and interaction effects.

**1. Introduction.** Predictive learning is a common application in data mining, machine learning and pattern recognition. The purpose is to predict the unknown value of an attribute $y$ of a system under study, using the known joint values of other attributes $\mathbf{x} = (x_1, x_2, \ldots, x_n)$ associated with that system. The prediction takes the form $\hat{y} = F(\mathbf{x})$, where the function $F(\mathbf{x})$ maps a set of joint values of the "input" variables $\mathbf{x}$ to a value $\hat{y}$ for the "output" variable $y$. The goal is to produce an accurate mapping. Lack of accuracy is defined by the prediction "risk"

$$R(F) = E_{\mathbf{x}y} L(y, F(\mathbf{x})), \tag{1}$$

Received May 2007; revised October 2007.
[1]Supported in part by the Department of Energy under contract DE-AC02-76SF00515 and NSF Grant DMS-97-64431.
[2]Supported in part by NSF Grant DMS-97-64431.

where $L(y, \hat{y})$ represents a loss or cost for predicting a value $\hat{y}$ when the actual value is $y$, and the expected value is over the joint distribution of all variables $(\mathbf{x}, y)$ for the data to be predicted. Using this definition, the optimal mapping ("target") function is given by

$$(2) \qquad F^*(\mathbf{x}) = \arg \min_{F(\mathbf{x})} E_{\mathbf{x}y} L(y, F(\mathbf{x})).$$

With predictive learning, one is given a "training" sample of previously solved cases $\{\mathbf{x}_i, y_i\}_1^N$ where the joint values of all variables have been determined. An approximation $F(\mathbf{x})$ to $F^*(\mathbf{x})$ is derived by applying a learning procedure to these data.

**2. Ensemble learning.** Learning ensembles have emerged as being among the most powerful learning methods [see Breiman (1996, 2001), Freund and Schapire (1996), Friedman (2001)]. Their structural model takes the form

$$(3) \qquad F(\mathbf{x}) = a_0 + \sum_{m=1}^{M} a_m f_m(\mathbf{x}),$$

where $M$ is the size of the ensemble and each ensemble member ("base learner") $f_m(\mathbf{x})$ is a different function of the input variables $\mathbf{x}$ derived from the training data. Ensemble predictions $F(\mathbf{x})$ are taken to be a linear combination of the predictions of each of the ensemble members, with $\{a_m\}_0^M$ being the corresponding parameters specifying the particular linear combination. Ensemble methods differ in choice of particular base learners (function class), how they are derived from the data, and the prescription for obtaining the linear combination parameters $\{a_m\}_0^M$.

The approach taken here is based on the importance sampled learning ensemble (ISLE) methodology described in Friedman and Popescu (2003). Given a set of base learners $\{f_m(\mathbf{x})\}_1^M$, the parameters of the linear combination are obtained by a regularized linear regression on the training data $\{\mathbf{x}_i, y_i\}_1^N$,

$$(4) \quad \{\hat{a}_m\}_0^M = \arg \min_{\{a_m\}_0^M} \sum_{i=1}^{N} L\left(y_i, a_0 + \sum_{m=1}^{M} a_m f_m(\mathbf{x}_i)\right) + \lambda \cdot \sum_{m=1}^{M} |a_m|.$$

The first term in (4) measures the prediction risk (1) on the training sample, and the second (regularization) term penalizes large values for the coefficients of the base learners. The influence of this penalty is regulated by the value of $\lambda \geq 0$. It is well known that for this ("lasso") penalty, larger values of $\lambda$ produce more overall shrinkage as well as increased dispersion among the values $\{|\hat{a}_m|\}_1^M$, often with many being set to zero [see Tibshirani (1996), Donoho et al. (1995)]. Its value is taken to be that which minimizes an estimate of future prediction risk (1) based on a separate sample not used in training, or by full (multi-fold) cross-validation. Fast algorithms for solving (4) for all values of $\lambda \geq 0$, using a variety of loss functions $L(y, \hat{y})$, are presented in Friedman and Popescu (2004).

The base learners $\{f_m(\mathbf{x})\}_1^M$ used in (3) and (4) to characterize the ensemble are randomly generated using the perturbation sampling technique described in Friedman and Popescu (2003). Each one is taken to be a simple function of the predictor variables characterized by a set of parameters $\mathbf{p} = (p_1, p_2, \ldots)$. That is,

$$(5) \qquad f_m(\mathbf{x}) = f(\mathbf{x}; \mathbf{p}_m),$$

where $\mathbf{p}_m$ represents a specific set of joint parameter values indexing a specific function $f_m(\mathbf{x})$ from the parameterized class $f(\mathbf{x}; \mathbf{p})$. Particular choices for such parameterized function classes are discussed below.

Given a function class, the individual members of the ensemble are generated using the prescription presented in Friedman and Popescu (2003) and shown in Algorithm 1.

ALGORITHM 1 (Ensemble generation).
1  $F_0(\mathbf{x}) = \arg\min_c \sum_{i=1}^N L(y_i, c)$
2  For $m = 1$ to $M$ {
3      $\mathbf{p}_m = \arg\min_{\mathbf{p}} \sum_{i \in S_m(\eta)} L(y_i, F_{m-1}(\mathbf{x}_i) + f(\mathbf{x}_i; \mathbf{p}))$
4      $f_m(\mathbf{x}) = f(\mathbf{x}; \mathbf{p}_m)$
5      $F_m(\mathbf{x}) = F_{m-1}(\mathbf{x}) + \nu \cdot f_m(\mathbf{x})$
6  }
7  ensemble $= \{f_m(\mathbf{x})\}_1^M$

In line 3, $S_m(\eta)$ represents a different subsample of size $\eta < N$ randomly drawn without replacement from the original training data, $S_m(\eta) \subset \{\mathbf{x}_i, y_i\}_1^N$. As discussed in Friedman and Popescu (2003), smaller values of $\eta$ encourage increased dispersion (less correlation) among the ensemble members $\{f_m(\mathbf{x})\}_1^M$ by training them on more diverse subsamples. Smaller values also reduce computation by a factor of $N/\eta$.

At each step $m$, the "memory" function

$$F_{m-1}(\mathbf{x}) = F_0(\mathbf{x}) + \nu \cdot \sum_{k=1}^{m-1} f_k(\mathbf{x})$$

contains partial information concerning the previously induced ensemble members $\{f_k(\mathbf{x})\}_1^{m-1}$ as controlled by the value of the "shrinkage" parameter $0 \leq \nu \leq 1$. At one extreme, setting $\nu = 0$ causes each base learner $f_m(\mathbf{x})$ to be generated without reference to those previously induced, whereas the other extreme $\nu = 1$ maximizes their influence. Intermediate values $0 < \nu < 1$ vary the degree to which previously chosen base learners effect the generation of each successive one in the sequence.

Several popular ensemble methods represent special cases of Algorithm 1. A "bagged" ensemble [Breiman (1996)] is obtained by using squared-error loss, $L(y, \hat{y}) = (y - \hat{y})^2$, and setting $\nu = 0$, and $\eta = N/2$ or, equivalently, choosing $S_m$ (line 3) to be a bootstrap sample [Friedman and Hall (2007)]. Random forests [Breiman (2001)] introduce increased ensemble dispersion by additionally randomizing the algorithm ("arg min," line 3) used to solve for the ensemble members (large decision trees). In both cases the coefficients in (3) are set to $a_0 = \bar{y}$, $\{a_m = 1/M\}_1^M$ so that predictions are a simple average of those of the ensemble members.

AdaBoost [Freund and Schapire (1996)] uses exponential loss, $L(y, \hat{y}) = \exp(-y \cdot \hat{y})$ for $y \in \{-1, 1\}$, and is equivalent to setting $\nu = 1$ and $\eta = N$ in Algorithm 1. Predictions are taken to be the sign of the final memory function $F_M(\mathbf{x})$. MART [Friedman (2001)] uses a variety of loss criteria $L(y, \hat{y})$ for arbitrary $y$, and in default mode sets $\nu = 0.1$ and $\eta = N/2$. Predictions are given by $F_M(\mathbf{x})$.

Friedman and Popescu (2003) experimented with a variety of joint $(\nu, \eta)$ values for generating ensembles of small decision trees, followed by (4) to estimate the linear combination parameters. Their empirical results indicated that small but nonzero values of $\nu$ ($\nu \simeq 0.01$) performed best in this context. Results were seen to be fairly insensitive to the value chosen for $\eta$, provided it was small ($\eta \lesssim N/2$) and grew less rapidly than the total sample size $N$ ($\eta \sim \sqrt{N}$) as $N$ becomes large ($N \gtrsim 500$).

Although, in principle, most of these procedures can be used with other base learners, they have almost exclusively been applied with decision trees [Breiman et al. (1983), Quinlan (1993)].

**3. Rule based ensembles.** The base learners considered here are simple rules. Let $S_j$ be the set of all possible values for input variable $x_j$, $x_j \in S_j$, and $s_{jm}$ be a specified subset of those values, $s_{jm} \subseteq S_j$. Then each base learner takes the form of a conjunctive rule

$$(6) \qquad r_m(\mathbf{x}) = \prod_{j=1}^{n} I(x_j \in s_{jm}),$$

where $I(\cdot)$ is an indicator of the truth of its argument. Each such base learner assumes two values $r_m(\mathbf{x}) \in \{0, 1\}$. It is nonzero when all of the input variables realize values that are simultaneously within their respective subsets $\{x_j \in s_{jm}\}_1^n$. For variables that assume orderable values, the subsets are taken to be contiguous intervals

$$s_{jm} = (t_{jm}, u_{jm}]$$

defined by a lower and upper limit, $t_{jm} < x_j \leq u_{jm}$. For categorical variables assuming unorderable values (names), the subsets are explicitly enumerated.

Such rules (6) can be regarded as parameterized functions of $\mathbf{x}$ (5), where the parameters $\mathbf{p}_m$ are the quantities that define the respective subsets $\{s_{jm}\}$.

Note that for the case in which the subset of values $s_{jm}$ (real or categorical) appearing in a factor of (6) is in fact the entire set $s_{jm} = S_j$, the corresponding factor can be omitted from the product. In this case the rule (6) can be expressed in the simpler form

$$(7) \qquad r_m(\mathbf{x}) = \prod_{s_{jm} \neq S_j} I(x_j \in s_{jm}).$$

The particular input variables $x_j$ for which $s_{jm} \neq S_j$ are said to be those that "define" the rule $r_m(\mathbf{x})$. For purposes of interpretation, it is desirable that the ensemble be comprised of "simple" rules each defined by a small number of variables. As an example, the rule

$$r_m(\mathbf{x}) = \begin{cases} I(18 \leq \text{age} < 34) \\ \cdot I(\text{marital status} \in \{\text{single, living together-not married}\}) \\ \cdot I(\text{householder status} = \text{rent}) \end{cases}$$

is defined by three variables, and a nonzero value increases the odds of frequenting bars and night clubs.

3.1. *Rule generation.* One way to attempt to generate a rule ensemble is to let the base learner $f(\mathbf{x}; \mathbf{p})$ appearing in Algorithm 1 take the form of a rule (6) and then try to solve the optimization problem on line 3 for the respective variable subsets $\{s_{jm}\}$. Such a (combinatorial) optimization is generally infeasible for more that a few predictor variables, although fast approximate algorithms can be employed [Cohen and Singer (1999), Weiss and Indurkhya (2000)]. The approach used here is to view a decision tree as defining a collection of rules and take advantage of existing fast algorithms for producing decision tree ensembles. That is, decision trees are used as the base learner $f(\mathbf{x}; \mathbf{p})$ in Algorithm 1. Each node (interior and terminal) of each resulting tree $f_m(\mathbf{x})$ produces a rule of the form (7).

This is illustrated in Figure 1 which shows a typical decision tree with five terminal nodes that could result from using a decision tree algorithm in conjunction with Algorithm 1. Associated with each interior node is one of the input variables $x_j$. For variables that realize orderable values, a particular value of that variable ("split point") is also associated with the node. For variables that assume unorderable categorical values, a specified subset of those values replaces the split point. For the tree displayed in Figure 1, nodes 0 and 4 are associated with orderable variable $x_{14}$ with split points $u$ and $t$ respectively, node 1 is associated with categorical variable $x_{32}$ with subset values $\{a, b, c\}$, and node 2 is associated with categorical variable $x_7$ with the single value $\{z\}$.

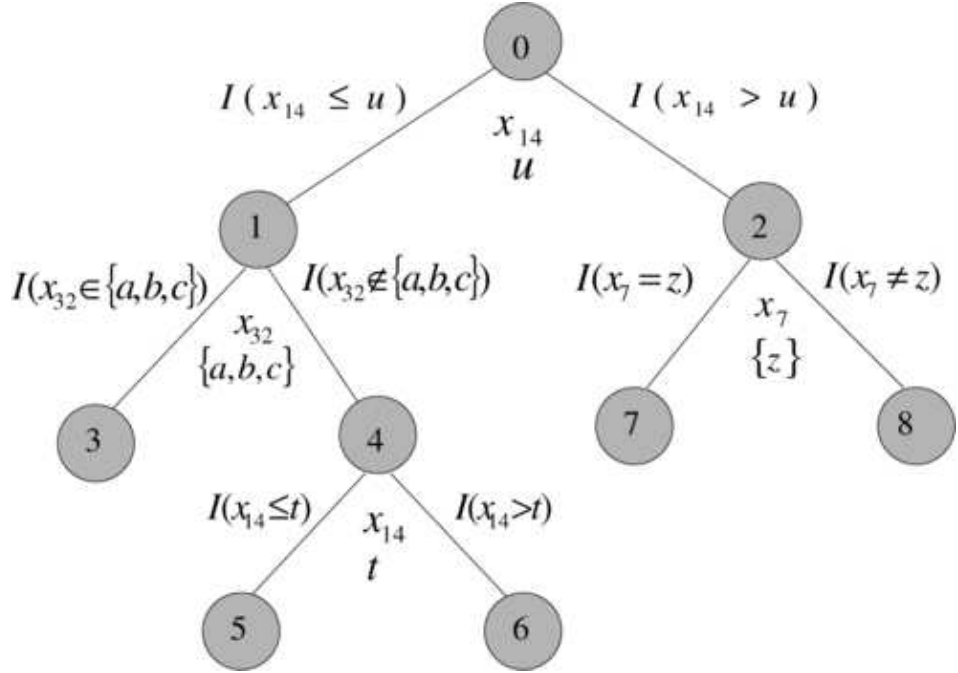

Fig. 1. *A decision tree. The rule corresponding to each node is given by the product of the indicator functions associated with all of the edges on the path from the root to that node.*

Each edge of the tree connecting a "parent" node to one of its two "daughter" nodes represents a factor in (7) contributing to the rules corresponding to all descendent nodes of the parent. These factors are shown in Figure 1 for each such edge. The rule corresponding to any node in the tree is given by the product of the factors associated with all of the edges on the path from the root to that node. Note that there is no rule corresponding to the root node. As examples, in Figure 1 the rules corresponding to nodes 1, 4, 6, and 7 are respectively:

$$r_1(\mathbf{x}) = I(x_{14} \leq u),$$
$$r_4(\mathbf{x}) = I(x_{14} \leq u) \cdot I(x_{32} \notin \{a,b,c\}),$$
$$r_6(\mathbf{x}) = I(t < x_{14} \leq u) \cdot I(x_{32} \notin \{a,b,c\}),$$
$$r_7(\mathbf{x}) = I(x_{14} > u) \cdot I(x_7 = z).$$

3.2. *Rule fitting.* The collection of all such rules derived from all of the trees $\{f_m(\mathbf{x})\}_1^M$ produced by Algorithm 1 constitute the rule ensem-ble $\{r_k(\mathbf{x})\}_1^K$. The total number of rules is

$$K = \sum_{m=1}^{M} 2(t_m - 1), \tag{8}$$

where $t_m$ is the number of terminal nodes for the $m$th tree. The predictive model is

$$F(\mathbf{x}) = \hat{a}_0 + \sum_{k=1}^{K} \hat{a}_k r_k(\mathbf{x}), \tag{9}$$

with

$$\{\hat{a}_k\}_0^K = \arg \min_{\{a_k\}_0^K} \sum_{i=1}^{N} L\left(y_i, a_0 + \sum_{k=1}^{K} a_k r_k(\mathbf{x}_i)\right) + \lambda \cdot \sum_{k=1}^{K} |a_k|. \tag{10}$$

Fast algorithms for solving (10) for all values of $\lambda \geq 0$, and procedures for choosing a value for $\lambda$, are discussed in Friedman and Popescu (2004).

The solution to (10) for $\lambda > 0$ is not equivariant under different scaling transformations applied to each of the predicting rules $r_k(\mathbf{x})$. Increasing the scale of a rule by $r_k(\mathbf{x}) \leftarrow b_k \cdot r_k(\mathbf{x})$ ($b_k > 1$) and decreasing its corresponding coefficient $a_k \leftarrow a_k/b_k$ produces the same loss in the first term of (10), but reduces its contribution to the second penalty term. Therefore, the coefficients of rules with larger scales are penalized less than those with smaller scales. The scale of a rule is characterized by its standard deviation

$$t_k = \sqrt{s_k(1 - s_k)}, \tag{11}$$

where $s_k$ is its support on the training data

$$s_k = \frac{1}{N} \sum_{i=1}^{N} r_k(\mathbf{x}_i). \tag{12}$$

A common practice is to give all predictors equal a priori influence, for example, by replacing each rule by a normalized version $r_k(\mathbf{x}) \leftarrow r_k(\mathbf{x})/t_k$ in (10). The strategy applied here is to use the original *unnormalized* rules in (10). This places increased penalty on coefficients of rules with very small support $s_k \simeq 0$ and on those with very large support $s_k \simeq 1$. The overall effect is to reduce the variance of the estimated model (9) since rules with such small support, or the complement of those with such large support, are each defined by a correspondingly small number of training observations.

3.3. *Tree size.* As seen in Figure 1 the size of each tree, as characterized by the number of its terminal nodes, along with the tree topology, determines the maximum number of factors appearing in the rules (7) derived from that tree. The topology of each individual tree is determined by the data. However, larger trees generally allow more complex rules to be produced in terms of the number of variables (factors) that define them. For example, the smallest trees with only two terminal nodes ("stumps") generate rules limited to one factor in (7), whereas an $L$ terminal node tree can, in principle, generate rules involving up to $L-1$ factors. Thus, controlling tree size directly controls maximum complexity, and indirectly the average complexity, of the rules that comprise the ensemble.

Controlling tree size, and thereby average rule complexity, also influences the type of target functions (2) that are most easily approximated by the ensemble. In order to capture interaction effects involving $l$ variables, the ensemble must include rules with $l$ or more factors. Thus, targets that involve strong high order interaction effects require larger trees than those that are dominately influenced by main effects and/or low order interactions. On the other hand, for a given size $K$ (8), ensembles comprised of a large fraction of high order interaction rules will necessarily involve fewer of lower order that are best able to capture main and low order interaction effects. Therefore, larger trees can be counter productive for targets of this latter type. The best tree size is thus governed by the nature of the (unknown) target function.

The strategy used here is to produce an ensemble of trees of varying sizes from which to extract the rules by letting the number of terminal nodes $t_m$ of each tree be a random variable

$$t_m = 2 + fl(\gamma).$$

Here $\gamma$ is randomly drawn from an exponential distribution with probability

$$\text{(13)} \qquad \Pr(\gamma) = \exp(-\gamma/(\bar{L}-2))/(\bar{L}-2),$$

and $fl(\gamma)$ is the largest integer less than or equal to $\gamma$. The quantity $\bar{L} \geq 2$ represents the average number of terminal nodes for trees in the ensemble. For $\bar{L} = 2$, the entire ensemble will be composed of rules each involving only one of the input variables and thereby capture main effects only. Larger values produce trees of varying size $t_m$, mostly with $t_m \leq \bar{L}$, but many with $t_m > \bar{L}$ and some with $t_m \gg \bar{L}$ producing some rules capable of capturing high order interactions, if present. The fitting procedure (10) can then attempt to select those rules most relevant for prediction. The use of an exponential distribution (13) counters the tendency of trees (of a given size) to produce more rules involving a larger number of factors owing to their hierarchical (binary tree) topology. The overall result is a more evenly distributed ensemble in terms of the complexity of its rules.

The average tree size $\bar{L}$ is a "meta"-parameter of the procedure that controls the distribution of the complexity of the rules $\{r_k(\mathbf{x})\}_1^K$ comprising the ensemble. A choice for its value can be based on prior suspicions concerning the nature of the target $F^*(\mathbf{x})$, or one can experiment with several values using an estimate of future predictive accuracy based on an independent sample or cross-validation. Also, examination of the actual rules chosen for prediction in (10) can suggest potential modifications.

3.4. *Loss functions.* Any predictive learning method involves the specification of a loss function $L(y, F)$ that characterizes the loss or cost of predicting an outcome or response value $F$ when the actual value is $y$. As described in Friedman and Popescu (2003, 2004), the ensemble procedures presented here can be implemented with a variety of different loss criteria. Specific choices can have a substantial effect on predictive models estimated from data, and are appropriate in different settings. For example, if the deviations from the target $F^*(\mathbf{x})$ (2) follow a (homoskedastic) Gaussian distribution

$$(14) \qquad y_i \sim N(F^*(\mathbf{x}_i), \sigma^2),$$

then squared-error loss

$$(15) \qquad L(y, F) = (y - F)^2$$

is most appropriate.

For other distributions of a numeric outcome variable $y$, and especially in the presence of outliers, the Huber (1964) loss

$$(16) \qquad L(y, F) = \begin{cases} (y - F)^2/2, & |y - F| < \delta, \\ \delta(|y - F| - \delta/2), & |y - F| \geq \delta, \end{cases}$$

provides increased robustness, while sacrificing very little accuracy in situations characterized by (14) [see Friedman and Popescu (2004)]. It is a compromise between squared-error loss (15) and absolute deviation loss $L(y, F) = |y - F|$. The value of the "transition" point $\delta$ differentiates the residuals that are treated as outliers being subject to absolute loss, from the other residuals subject to squared-error loss. Its value is taken to be the $\alpha$th quantile of the data absolute residuals $\{|y_i - F(\mathbf{x}_i)|\}_1^N$, where the value of $\alpha$ controls the degree of robustness (break down) of the procedure; smaller values produce more robustness. For the simulated regression problems illustrated in the following squared-error loss, (15) is used, whereas for the real data example Huber loss, (16) with $\alpha = 0.9$ was employed to guard against potential outliers.

For binary classification $y \in \{-1, 1\}$, a variety of loss criteria have been proposed [see Hastie, Tibshirani and Friedman (2001)]. Here we use the squared-error ramp loss

$$(17) \qquad L(y, F) = [y - \min(-1, \max(1, F))]^2$$

introduced and studied in Friedman and Popescu (2001, 2004). It was shown to produce comparable performance to other commonly used loss criteria, but with increased robustness against mislabeled training cases.

**4. Accuracy.** An important property of any learning method is accuracy as characterized by its prediction risk (1). As noted in Section 2, decision tree ensembles are among the most competitive methods. Friedman and Popescu (2001) compared the performance of several decision tree ensemble methods in a simulation setting. These included bagging, random forests, boosting, and a variety of ISLEs using Algorithm 1 to construct the tree ensembles with various joint values for $\eta$ and $\nu$, followed by (4) to estimate the linear combination parameters. Here we compare the performance of rule based ensembles discussed in Section 3 to best performing tree based ensembles studied there.

The simulation consisted of 100 data sets, each with $N = 10000$ observations and $n = 40$ input variables. Each data set was generated from the model

$$\tag{18} \{y_i = F^*(\mathbf{x}_i) + \varepsilon_i\}_1^N,$$

with $F^*(\mathbf{x})$ being a different target function for each data set. These 100 target functions were themselves each randomly generated so as to produce a wide variety of different targets in terms of their dependence on the input variables $\mathbf{x}$. Details concerning this random function generator are presented in Friedman (2001) and also in Friedman and Popescu (2003). The input variables were randomly generated according to a standard Gaussian distribution $x_j \sim N(0,1)$. The irreducible error $\varepsilon$ was also randomly generated from a Gaussian, $\varepsilon \sim N(0, \sigma^2)$, with $\sigma^2 = \text{Var}_{\mathbf{x}} F^*(\mathbf{x})$ to produce a one-to-one signal-to-noise ratio. In addition to regression, data for binary classification was produced by thresholding the response values for each data set at their respective medians

$$\tag{19} \{\tilde{y}_i = \text{sign}(y_i - \text{median}(\{y_k\}_1^N))\}_1^N.$$

The resulting optimal decision boundaries for each data set are quite different and fairly complex.

Here we present a comparison of four methods. The first "MART" [Friedman (2001)] is a popular tree boosting method. The second "ISLE" is the best performing tree ensemble considered in Friedman and Popescu (2003) as averaged over these 100 data sets. It uses Algorithm 1 to generate the trees with $\eta = N/5$ and $\nu = 0.01$, followed by (4) to estimate the linear combination parameters. In both cases the ensembles consisted of 500 six-terminal node trees. The third method "RuleFit" here uses exactly the same tree ensemble produced by ISLE to facilitate comparison, but then extracts the ten rules associated with each of the trees as described in Section 3.1. The resulting collection of $K = 5000$ rules (8) is then used to form the predictive model (9), (10). The last method RuleFit 200 uses the same procedure except that only the first 200 trees are used to extract $K = 2000$ rules for the final model. Although a large number of rules are used to fit the model in (10), typically only a small fraction ($\sim 10\%$) have nonzero solution coefficient values and are thus required for prediction in (9).

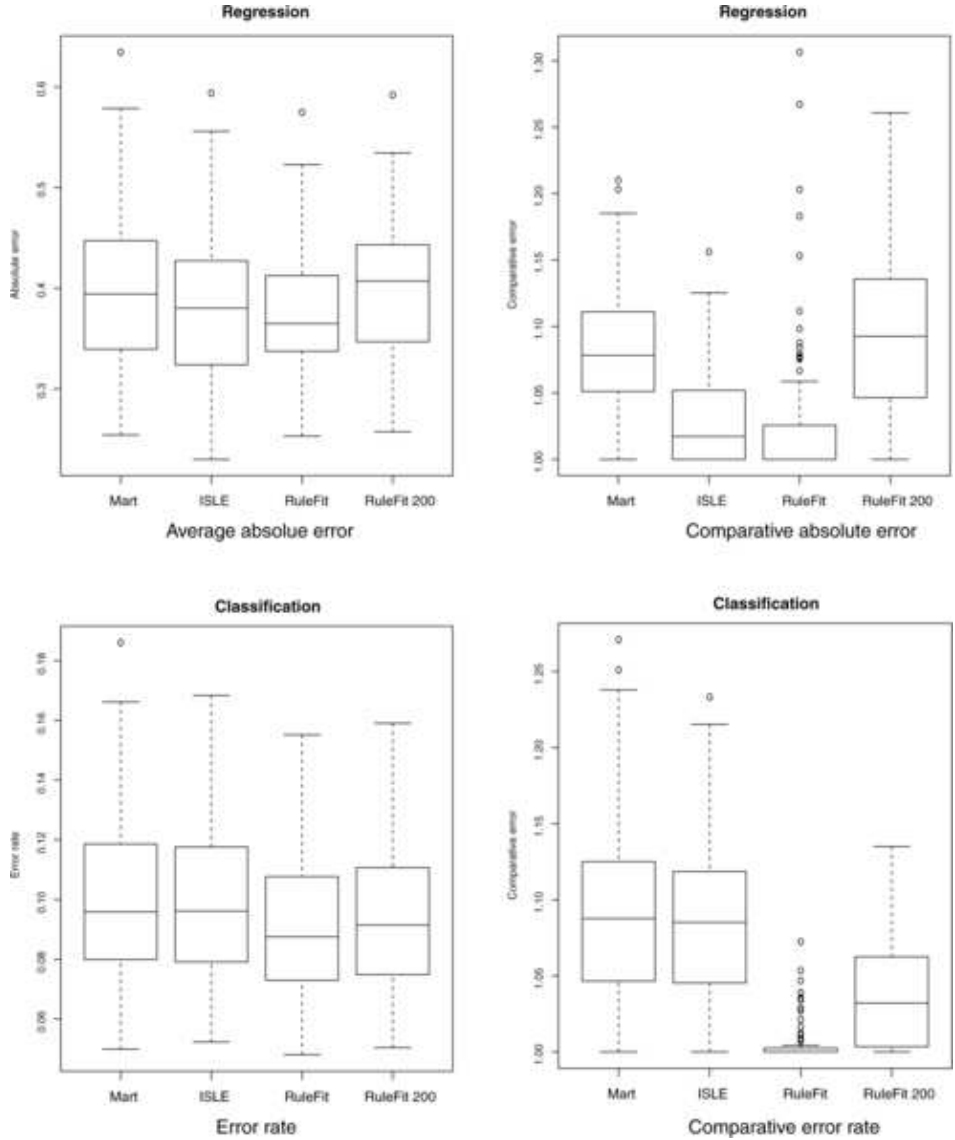

Fig. 2. *Inaccuracy comparisons between tree ensemble methods (Mart, ISLE) and rule based ensembles (RuleFit, RuleFit 200).*

The upper left panel of Figure 2 shows the distributions (box plots) of the scaled absolute error

$$(20) \qquad e_{jl} = \frac{E_{\mathbf{x}}[|F_l^*(\mathbf{x}) - F_{jl}(\mathbf{x})|]}{E_{\mathbf{x}}[|F_l^*(\mathbf{x}) - \text{median } F_l^*(\mathbf{x})|]}, \qquad l = 1, 100,$$

over the 100 regression data sets for each of the four methods. Here $F_l^*(\mathbf{x})$ is the true target function for the $l$th data set, and $F_{jl}(\mathbf{x})$ is the corresponding estimate for the $j$th method ($j = 1, 4$). One sees that these 100 target functions represent a wide range of difficulty for all methods and that on average RuleFit provides slightly better performance. Using rules based on only 200 trees is still competitive with the 500 tree MART ensemble, but somewhat inferior to the 500 tree ISLE on these typically fairly complex target functions.

The upper right panel of Figure 2 shows the corresponding distributions of the *comparative* absolute error defined by

$$(21) \qquad c_{jl} = e_{jl}/\min\{e_{kl}\}_{k=1}^4.$$

This quantity facilitates individual comparisons by using the error of the best method for each data set to calibrate the difficulty of each respective problem. The best method $j^* = \arg\min_j\{e_{jl}\}_{j=1}^4$ for each data set receives the value $c_{j^*l} = 1$, and the others larger values in proportion to their average error (20) on that data set. Here one sees that RuleFit based on 500 trees yields the best performance, or close to it, on nearly all of the 100 data sets. There are a few ($\sim 5$) for which one of the other methods was distinctly better. Of course, there are many for which the converse holds.

The lower panels of Figure 2 show the corresponding results for the classification (19). Here lack of performance is measured in terms of error rate

$$(22) \qquad e_{jl} = E_{\mathbf{x}} I[\tilde{y} \neq \text{sign}(F_{jl}(\mathbf{x}))].$$

Again, these 100 classification problems present varying degrees of difficulty for all methods with error rates ranging by roughly a factor of three. Both rule based methods exhibit slightly superior average classification performance to the tree based ensembles. This is especially reflected in the corresponding comparative error rates (21), (22) shown in the lower right panel where RuleFit based on 500 trees was the best on almost every data set, and even RuleFit 200 was substantially better than either of the tree based ensembles with 500 trees.

The results presented in Figure 2 suggest that the rule based approach to ensemble learning described in Section 3 produces accuracy comparable to that based on decision trees. Other tree based ensemble methods including bagging and random forests were compared to those presented here (MART, ISLE) in Friedman and Popescu (2003), and seen to exhibit somewhat lower accuracy over these 100 regression and classification data sets. Thus, rule based ensembles appear to be competitive in accuracy with the best tree based ensembles.

**5. Linear basis functions.** With ensemble learning there is no requirement that the basis functions $\{f_m(\mathbf{x})\}_1^M$ in (3) and (4) must be generated from the same parametric base learner (5). Other basis functions can be included, either generated from another parametric family using Algorithm 1, or by some other means. For increased accuracy, the different families should be chosen to complement each other in that each is capable of closely approximating target functions (2) for which the others have difficulty. For the purpose of interpretation, each such family should also produce easily understandable basis functions.

Among the most difficult functions for rule (and tree) based ensembles to approximate are linear functions

$$(23) \qquad F^*(\mathbf{x}) = b_0 + \sum_{j=1}^{n} b_j x_j,$$

for which a substantial number of the coefficients $b_j$ have relatively large absolute values. Such targets can require a large number of rules for accurate approximation. Especially if the training sample is not large and/or the signal-to-noise ratio is small, it may not be possible to reliably estimate models with large enough rule sets. Also, models with many roughly equally contributing rules are more difficult to interpret.

These considerations suggest that both accuracy and interpretability might be improved by including the original variables $\{x_j\}_1^n$ as additional basis functions in (9) and (10) to complement the rule ensemble. In the interest of robustness against input variable outliers we use the "Winsorized" versions

$$(24) \qquad l_j(x_j) = \min(\delta_j^+, \max(\delta_j^-, x_j)),$$

where $\delta_j^-$ and $\delta_j^+$ are respectively the $\beta$ and $(1-\beta)$ quantiles of the data distribution $\{x_{ij}\}_{i=1}^N$ for each variable $x_j$. The value chosen for $\beta$ reflects ones, prior suspicions concerning the fraction of such outliers. Depending on the nature of the data, small values ($\beta \simeq 0.025$) are generally sufficient.

With these additions, the predictive model (9) becomes

$$(25) \qquad F(\mathbf{x}) = \hat{a}_0 + \sum_{k=1}^{K} \hat{a}_k r_k(\mathbf{x}) + \sum_{j=1}^{n} \hat{b}_j l_j(x_j),$$

with

$$(26) \quad (\{\hat{a}_k\}_0^K, \{\hat{b}_j\}_1^n) = \arg\min_{\{a_k\}_0^K, \{b_j\}_1^n} \sum_{i=1}^{N} L\left(y_i, a_0 + \sum_{k=1}^{K} a_k r_k(\mathbf{x}_i) + \sum_{j=1}^{n} b_j l_j(x_{ij})\right)$$
$$+ \lambda \cdot \left(\sum_{k=1}^{K} |a_k| + \sum_{j=1}^{n} |b_j|\right).$$

In order to give each linear term (24) the same a priori influence as a typical rule, its normalized version

$$l_j(x_j) \leftarrow 0.4 \cdot l_j(x_j)/std(l_j(x_j))$$

is used in (26), and then the corresponding solution coefficients $\{\hat{b}_j\}_1^n$ (and intercept $\hat{a}_0$) are transformed to reference the original $l_j(x_j)$ (24). Here $std(l_j(x_j))$ is the standard deviation of $l_j(x_j)$ over the training data and $0.4$ is the average standard deviation (11) of rules with a uniform support distribution $s_k \sim U(0,1)$.

Owing to the selective nature of the lasso penalty in (26), many of the rule coefficient estimates $\hat{a}_k$ as well as those $\hat{b}_j$ of the less influential linear variables will often have zero values, and thus need not appear in the final predictive model (25).

5.1. *Illustration.* To illustrate the potential benefit of including the original variables (24) as part of the ensemble, we consider simulated data generated from the model

$$(27) \qquad \left\{ y_i = 10 \cdot \prod_{j=1}^{5} e^{-2x_{ij}^2} + \sum_{j=6}^{35} x_{ij} + \varepsilon_i \right\}_{i=1}^{N},$$

with $N = 10000$ observations and $n = 100$ input variables, of which 65 have no influence on the response $y$. There is a strong nonlinear dependence on the first five input variables and a linear dependence of equal strength on 30 others. All input variables were randomly generated from a uniform distribution, $x_{ij} \sim U(0,1)$, and the irreducible noise $\varepsilon_i$ was generated from a Gaussian distribution, $\varepsilon_i \sim N(0, \sigma^2)$, with $\sigma$ chosen to produce a two-to-one signal-to-noise ratio.

Figure 3 shows the distribution (box plots) of the scaled absolute error (20) over 100 data sets randomly generated according to the above prescription, for three ensembles. The first "linear" involves no rules; only the $n = 100$ linear variables (24) comprise the ensemble. The second ensemble "rules" consists of $K = 2000$ rules generated as described in Section 3. The third ensemble "both" is the union of the first two; it includes the 100 linear variables and the 2000 rules. As seen in Figure 3, the purely linear model exhibits relatively poor performance; it has trouble capturing the highly nonlinear dependence on the first five input variables (27). The ensemble based only on rules provides somewhat improved performance by being better able to approximate the nonlinearity while crudely approximating the linear dependence by piecewise constants. The ensemble based on both linear variables and rules here provides the highest accuracy. The selection effect of the lasso penalty in (26) tends to give high influence to the best rules for approximating the nonlinear dependencies as well as to the appropriate linear terms (24) for capturing the linear component in (27).

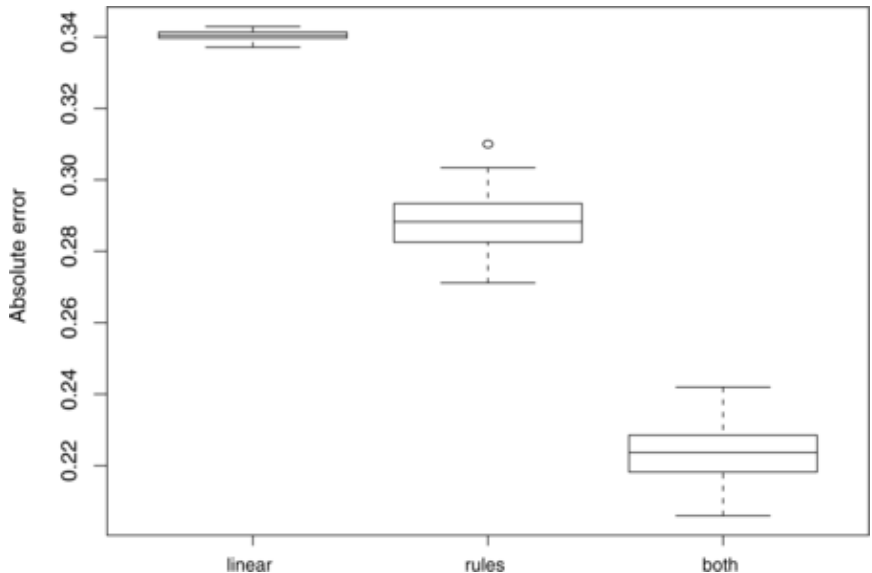

Fig. 3. *Average absolute error for linear model, rules only model, and combined rules and linear base learners.*

This example was constructed to especially illustrate the potential advantage of including linear basis functions as part of rule based ensembles. In many applications the corresponding improvement is less dramatic. For example, the target functions generated by the random function generator used in Section 4 tend to be very highly nonlinear [see Friedman (2001)] and the performance of the rule based ensembles including linear functions (not shown) was virtually identical to that based on rules alone as shown in Figure 2. Also in many applications the numeric variables realize a relatively small number of distinct values and the piecewise constant approximations based on relatively few rules are at less of a disadvantage at capturing linear dependencies. Including linear functions in the basis provides the greatest improvement in situations where there are a substantial number of relevant numeric variables, each realizing many distinct values, on which the target has an approximate linear dependence. However, even in settings unfavorable to linear basis functions, as in Section 4, their inclusion seldom degrades performance again owing to the selection effect of the lasso penalty in (26). In all the examples presented below the ensemble includes the linear functions (24) for all of the input variables as part of the basis.

**6. Rule based interpretation.** Rules of the form (7) represent easily understandable functions of the input variables $\mathbf{x}$, as do the linear functions (24). Although a large number of such functions participate in the initial ensemble, the fitting procedure (26) generally sets the vast majority ($\sim 80\%$ to $90\%$) of the corresponding coefficient estimates ($\{\hat{a}_k\}_1^K, \{\hat{b}_j\}_1^n$) to zero. As noted above, this selection property is a well-known aspect of the lasso penalty in (26). The remaining predictors [rules (7) or linear (24)] will have varying coefficient values depending on their estimated predictive relevance.

A commonly used measure of relevance or importance $I_k$ of any predictor in a linear model such as (25) is the absolute value of the coefficient of the corresponding standardized predictor. For rules, this becomes

$$(28) \qquad I_k = |\hat{a}_k| \cdot \sqrt{s_k(1-s_k)},$$

where $s_k$ is the rule support (12). For the linear predictors (24), the corresponding quantity is

$$(29) \qquad I_j = |\hat{b}_j| \cdot std(l_j(x_j)),$$

where $std(l_j(x_j))$ is the standard deviation of $l_j(x_j)$ over the data. Those predictors (rules or linear) with the largest values for (28) and (29) are the most influential for prediction based on the predictive equation (25). These can then be selected and examined for interpretation.

The importance measures (28) and (29) are global in that they reflect the average influence of each predictor over the distribution of all joint input variable values. A corresponding *local* measure of influence at *each* point $\mathbf{x}$ in that space is for rules (7)

$$(30) \qquad I_k(\mathbf{x}) = |\hat{a}_k| \cdot |r_k(\mathbf{x}) - s_k|,$$

and for the linear terms (24)

$$(31) \qquad I_j(x_j) = |\hat{b}_j| \cdot |l_j(x_j) - \bar{l}_j|,$$

where $\bar{l}_j$ is the mean of $l_j(x_j)$ over the training data. These quantities measure the (absolute) change in the prediction $F(\mathbf{x})$ when the corresponding predictor ($r_k(\mathbf{x})$ or $l_j(x_j)$) is removed from the predictive equation (25) and the intercept $\hat{a}_0$ is adjusted accordingly. That is, $\hat{a}_0 \leftarrow \hat{a}_0 - \hat{a}_k s_k$ for rules, and $\hat{a}_0 \leftarrow \hat{a}_0 - \hat{b}_j \bar{l}_j$ for linear predictors. Note that the average (root-mean-square) of (30) and (31) over all $\mathbf{x}$ values equates to the corresponding global measures (28) and (29).

For a given coefficient value $|\hat{a}_k|$, the importance (30) of the corresponding rule for a prediction at $\mathbf{x}$ depends on its value $r_k(\mathbf{x}) \in \{0,1\}$ at that point, as well as its global support (12). A rule is said to "fire" at a point $\mathbf{x}$ if $r_k(\mathbf{x}) = 1$. From (30) a rule that generally does not fire over the whole space ($s_k$ small) will have higher importance in regions where it does fire. Conversely, high support rules that usually fire will be correspondingly more important at points $\mathbf{x}$ where they do *not* fire, $r_k(\mathbf{x}) = 0$. This symmetry is a consequence of the fact that replacing a particular rule $r_k(\mathbf{x})$ by its complement $1 - r_k(\mathbf{x})$ produces an equivalent fitted linear model, so that either one should be assigned the same influence as reflected in (28) and (30).

The quantities (30) and (31) permit one to evaluate the relative influence of the respective predictors (rules or linear) for individual predictions $F(\mathbf{x})$ at $\mathbf{x}$. Those judged most influential can then be examined for interpreting that particular prediction. These quantities can also be averaged over selected subregions $S$ of the input variable space

$$(32) \qquad I_k(S) = \frac{1}{|S|} \sum_{\mathbf{x}_i \in S} I_k(\mathbf{x}_i); \qquad I_j(S) = \frac{1}{|S|} \sum_{\mathbf{x}_i \in S} I_j(x_{ij}),$$

where $|S|$ is the cardinality of $S$. For example, one might be interested in those predictors that most heavily influence relatively large predicted values

$$(33) \qquad S = \{\mathbf{x}_i | F(\mathbf{x}_i) \geq u\},$$

where the threshold $u$ might be a high quantile of the predictions $\{F(\mathbf{x}_i)\}_1^N$ over the data set. Similarly, one might define $S$ to be the set of lowest predicted values

$$(34) \qquad S = \{\mathbf{x}_i | F(\mathbf{x}_i) \leq t\},$$

with $t$ being a low quantile. In classification, $y \in \{-1, 1\}$, one might be interested in those rules that most heavily influence the predictions for each of the two respective classes. In this case $S = \{\mathbf{x}_i | y_i = 1\}$ or $S = \{\mathbf{x}_i | y_i = -1\}$ would be appropriate.

As with any linear model, the importance measures defined above are intended to estimate the influence of each individual predictor (rule or linear) after accounting for that of the others appearing in the ensemble. To the extent that the coefficient estimates are accurate, they will reflect the corresponding influence on the target function (2). These influence measures may or may not reflect the usefulness of individual predictors in the *absence* of others. For example, a predictor on which the target function (2) has no dependence at all may be useful if it is highly correlated with an important predictor, and the latter is removed from the ensemble. The influence measures used here are based on the joint contributions of all members of the ensemble.

**7. Input variable importance.** In predictive learning a descriptive statistic that is often of interest is the relative importance or relevance of the respective input variables $(x_1, x_2, \ldots, x_n)$ to the predictive model. For the models (25) considered here, the most relevant input variables are those that preferentially define the most influential predictors (rules or linear) appearing in the model. Input variables that frequently appear in important predictors are judged to be more relevant than those that tend to appear only in less influential predictors.

This concept can be captured by a measure of importance $J_l(\mathbf{x})$ of input variable $x_l$ at each individual prediction point $\mathbf{x}$ as

$$J_l(\mathbf{x}) = I_l(\mathbf{x}) + \sum_{x_l \in r_k} I_k(\mathbf{x})/m_k. \tag{35}$$

Here $I_l(\mathbf{x})$ is the importance (31) of the linear predictor (24) involving $x_l$, and the second term sums the importances of those rules (7) that contain $x_l$ ($x_l \in r_k$) each divided by the total number of input variables $m_k$ that define the rule. In this sense the input variables that define a rule equally share its importance, and rules with more variables do not receive exaggerated influence by virtue of appearing in multiple input variable importance measures.

The distribution of $\{J_l(\mathbf{x})\}_1^n$ (35) can be examined to ascertain the relative influence of the respective input variables locally at particular predictions $\mathbf{x}$. As with rules, these quantities can be averaged over selected subregions of the input variable space using (32), or over the whole space using (28) and (29), in place of the corresponding local measures in (35). Illustrations are provided in the data examples below.

**8. Interaction effects.** A function $F(\mathbf{x})$ is said to exhibit an interaction between two of its variables $x_j$ and $x_k$ if the difference in the value of $F(\mathbf{x})$ as a result of changing the value of $x_j$ depends on the value of $x_k$. For numeric variables, this can be expressed as

$$E_\mathbf{x}\left[\frac{\partial^2 F(\mathbf{x})}{\partial x_j \, \partial x_k}\right]^2 > 0$$

or by an analogous expression for categorical variables involving finite differences. If there is no interaction between these variables, the function $F(\mathbf{x})$ can be expressed as the sum of two functions, one that does not depend on $x_j$ and the other that is independent of $x_k$:

$$F(\mathbf{x}) = f_{\setminus j}(\mathbf{x}_{\setminus j}) + f_{\setminus k}(\mathbf{x}_{\setminus k}). \tag{36}$$

Here $\mathbf{x}_{\setminus j}$ and $\mathbf{x}_{\setminus k}$ respectively represent all variables except $x_j$ and $x_k$. If a given variable $x_j$ interacts with *none* of the other variables, then the function can be expressed as

$$F(\mathbf{x}) = f_j(x_j) + f_{\setminus j}(\mathbf{x}_{\setminus j}), \tag{37}$$

where the first term on the right is a function only of $x_j$ and the second is independent of $x_j$. In this case $F(\mathbf{x})$ is said to be "additive" in $x_j$.

A function $F(\mathbf{x})$ is said to have an interaction between three (numeric) variables $x_j$, $x_k$ and $x_l$ if

$$E_{\mathbf{x}}\left[\frac{\partial^3 F(\mathbf{x})}{\partial x_j\, \partial x_k\, \partial x_l}\right]^2 > 0,$$

again with an analogous expression involving finite differences for categorical variables. If there is no such three-variable interaction, $F(\mathbf{x})$ can be expressed as a sum of three functions, each independent of one of the three variables

(38) $$F(\mathbf{x}) = f_{\setminus j}(\mathbf{x}_{\setminus j}) + f_{\setminus k}(\mathbf{x}_{\setminus k}) + f_{\setminus l}(\mathbf{x}_{\setminus l}).$$

Here $\mathbf{x}_{\setminus j}$, $\mathbf{x}_{\setminus k}$ and $\mathbf{x}_{\setminus l}$ each respectively represent all of the variables except $x_j$, $x_k$ and $x_l$. Analogous expressions for the absence of even higher order interaction effects can be similarly defined.

Knowing which variables are involved in interactions with other variables, the identities of the other variables with which they interact, as well as the order and strength of the respective interaction effects can provide useful information about the predictive process as represented by the target function $F^*(\mathbf{x})$ (2). To the extent that the predictive model $F(\mathbf{x})$ (25), (26) accurately represents the target, one can infer these properties by studying its interaction effects.

As noted in Section 3.3, in order for the predictive model $F(\mathbf{x})$ (25) to capture an interaction among a specified subset of its variables, it is necessary that it contain rules (7) jointly involving all of the variables in the subset. This is, however, not a *sufficient* condition for the presence of such an interaction effect in $F(\mathbf{x})$. Different rules jointly involving these variables can combine to substantially reduce or possibly even eliminate various interaction effects between them as reflected in the overall model. Thus, the mere presence of rules involving multiple variables does not guarantee the existence of substantial interactions between the respective variables that define them. In order to uncover actual interaction effects, it is necessary to analyze the properties of the full predictive equation, not just individual components. Here we use the properties of partial dependence functions [Friedman (2001)] to study interaction effects in the predictive model.

8.1. *Partial dependence functions.* Given any subset $\mathbf{x}_s$ of the predictor variables indexed by $s \subset \{1, 2, \ldots, n\}$, the partial dependence of a function $F(\mathbf{x})$ on $\mathbf{x}_s$ is defined as

(39) $$F_s(\mathbf{x}_s) = E_{\mathbf{x}_{\setminus s}}[F(\mathbf{x}_s, \mathbf{x}_{\setminus s})],$$

where $\mathbf{x}_s$ is a prescribed set of joint values for the variables in the subset, and the expected value is over the marginal (joint) distribution of all variables $\mathbf{x}_{\setminus s}$ not represented in $\mathbf{x}_s$. Here $\mathbf{x} = (\mathbf{x}_s, \mathbf{x}_{\setminus s})$ is the entire variable set.

Partial dependence functions were used in Friedman (2001) to graphically examine the dependence of predictive models on low cardinality subsets of the variables, accounting for the averaged effects of the other variables. They can be estimated from data by

$$\hat{F}_s(\mathbf{x}_s) = \frac{1}{N} \sum_{i=1}^{N} F(\mathbf{x}_s, \mathbf{x}_{i\setminus s}), \tag{40}$$

where $\{\mathbf{x}_{i\setminus s}\}_1^N$ are the data values of $\mathbf{x}_{\setminus s}$. Here we use the properties of *centered* partial dependence functions to uncover and study interaction effects. In this section all partial dependence functions as well as the predictive model $F(\mathbf{x})$ (25) are considered to be centered to have a mean value of zero.

If two variables $x_j$ and $x_k$ do not interact, then from (36) the partial dependence of $F(\mathbf{x})$ on $\mathbf{x}_s = (x_j, x_k)$ can be decomposed into the sum of the respective partial dependences on each variable separately:

$$F_{jk}(x_j, x_k) = F_j(x_j) + F_k(x_k). \tag{41}$$

Furthermore, if a given variable $x_j$ does not interact with any other variable, then from (37) one has

$$F(\mathbf{x}) = F_j(x_j) + F_{\setminus j}(\mathbf{x}_{\setminus j}). \tag{42}$$

Here $F_{\setminus j}(\mathbf{x}_{\setminus j})$ is the partial dependence of $F(\mathbf{x})$ on all variables except $x_j$.

If variables $x_j$, $x_k$ and $x_l$ do not participate in a joint three-variable interaction, then from (38) the partial dependence of $F(\mathbf{x})$ on these three variables can be expressed in terms of the respective lower order partial dependencies as

$$\begin{aligned} F_{jkl}(x_j, x_k, x_l) &= F_{jk}(x_j, x_k) + F_{jl}(x_j, x_l) + F_{kl}(x_k, x_l) \\ &\quad - F_j(x_j) - F_k(x_k) - F_l(x_l). \end{aligned} \tag{43}$$

Analogous relationships can be derived for the absence of higher order interactions. These properties (41)–(43) of partial dependence functions are used to construct statistics to test for interaction effects of various types.

To test for the presence of an interaction between two specified variables $(x_j, x_k)$, the statistic

$$H_{jk}^2 = \sum_{i=1}^{N} [\hat{F}_{jk}(x_{ij}, x_{ik}) - \hat{F}_j(x_{ij}) - \hat{F}_k(x_{ik})]^2 \Big/ \sum_{i=1}^{N} \hat{F}_{jk}^2(x_{ij}, x_{ik}) \tag{44}$$

can be used based on (41) and the empirical estimates (40). It measures the fraction of variance of $\hat{F}_{jk}(x_j, x_k)$ not captured by $\hat{F}_j(x_j) + \hat{F}_k(x_k)$ over the data distribution. It will have a value of zero if the predictive model $F(\mathbf{x})$ (25), (26) exhibits no interaction between $x_j$ and $x_k$ and a correspondingly larger value for a stronger interaction effect between them. Similarly, a statistic for testing whether a specified variable $x_j$ interacts with *any* other variable would be from (42)

$$(45) \qquad H_j^2 = \sum_{i=1}^N [F(\mathbf{x}_i) - \hat{F}_j(x_{ij}) - \hat{F}_{\setminus j}(\mathbf{x}_{i\setminus j})]^2 \Big/ \sum_{i=1}^N F^2(\mathbf{x}_i).$$

This quantity will differ from zero to the extent that $x_j$ interacts with one or more other variables. By examining the values of $\{H_j\}_1^n$, one can identify those variables that participate in interaction effects. For each variable $x_j$ so identified, the statistics $\{H_{jk}\}_{k \neq j}$ (44) can be used to identify the variables that interact with $x_j$. Note that only those variables that are deemed globally relevant via (28), (29) and (35) need be considered for interaction effects. This is often a small subset of all $n$ predictor variables.

If a particular variable $x_j$ is seen to interact with more than one other variable using (44), it is of interest to ascertain the order of these interactions. That is, whether $x_j$ interacts separately with each of them or whether subsets of these variables jointly participate in higher order interactions. Let $x_k$ and $x_l$ be two variables that are identified as interacting with $x_j$. This could represent separate two-variable interactions between $(x_j, x_k)$ and $(x_j, x_l)$ only, or the additional presence of a three-variable interaction involving $(x_j, x_k, x_l)$. A statistic for testing these alternatives is from (43)

$$\begin{aligned}
H_{jkl}^2 = \sum_{i=1}^N [&\hat{F}_{jkl}(x_{ij}, x_{ik}, x_{il}) - \hat{F}_{jk}(x_{ij}, x_{ik}) \\
(46) \qquad & - \hat{F}_{jl}(x_{ij}, x_{il}) - \hat{F}_{kl}(x_{ik}, x_{il}) + \hat{F}_j(x_{ij}) \\
& + \hat{F}_k(x_{ik}) + \hat{F}_l(x_{il})]^2 \Big/ \sum_{i=1}^N \hat{F}_{jkl}^2(x_{ij}, x_{ik}, x_{il}).
\end{aligned}$$

This quantity tests for the presence of a joint three-variable interaction between $x_j$, $x_k$, and $x_l$ by measuring the fraction of variance of $\hat{F}_{jkl}(x_j, x_k, x_l)$ not explained by the lower order interaction effects among these variables. Analogous statistics testing for even higher order interactions can be derived, if desired.

By considering the *fraction* of unexplained variance, the statistics (44) and (46) test for the *presence* of the corresponding interaction effects in the predictive model $F(\mathbf{x})$ but do not necessarily reflect the importance of these effects to the overall variation of $F(\mathbf{x})$. It is possible for an interaction effect to be highly significant (see Section 8.3) but not very influential when compared to the other effects in the model. If for interpretational purposes one would like to uncover these as well as the more influential interactions, these statistics (44), (46) are appropriate. If it is desirable to ignore them so as to concentrate only on the highly influential interactions, then the statistics can be modified accordingly. Replacing the denominators in (44) and (46) with that in (45) will cause the resulting statistics to more closely reflect the importance of the corresponding interaction effects to the overall model $F(\mathbf{x})$.

8.2. *Spurious interactions.* The strategy outlined in the previous section is applied to the predictive model $F(\mathbf{x})$ (25) (26). As such, it will uncover interaction effects present in that model. However, interest is in interaction effects present in the target function $F^*(\mathbf{x})$ (2) representing the true underlying predictive relationships among the predictor variables $\mathbf{x}$. It is possible that even a highly accurate predictive model can contain substantial interaction effects that are not present in the target $F^*(\mathbf{x})$. These spurious interactions can occur when there is a high degree of collinearity among some (or all) of the predictor variables in the training data $\{\mathbf{x}_i\}_1^N$.

For example, if the target function exhibits a nonlinear additive dependence (37) on a variable $x_j$, this dependence on $x_j$ can be equivalently approximated by a corresponding additive contribution to the model involving that variable alone, or by incorporating interaction effects involving other variables highly correlated with it. Thus, it is not possible to easily distinguish between low and higher order interactions among subsets of variables that are highly correlated with each other. If interpretive value is to be placed on the presence of various interaction effects, then such spurious interactions should not be reported.

One way to discourage spurious interactions is to restrict their entry into the predictive model $F(\mathbf{x})$ (25), (26). Interactions enter the model through rules (7) involving more than one predictor variable. Such rules are derived from trees that have splits on different variables at nodes along the path from the root node to the nodes that define the respective rules (see Figure 1). Thus, one can discourage the entry of unneeded interaction effects by placing an incentive for fewer variables along each such path.

Trees are built in a top-down greedy fashion where the variable chosen for splitting at each node is the one that yields the maximal estimated improvement to tree predictions as a result of the split. The improvement $Z_j$ by potentially splitting the node on variable $x_j$ is estimated for all variables, and the one

$$j^* = \arg \max_{1 \leq j \leq n} Z_j$$

is chosen for splitting the node in question. Spurious interactions can be discouraged by modifying this splitting strategy so as to place an *incentive* for repeated splits on the same variable. Specifically,

$$j^* = \arg \max_{1 \leq j \leq n} \kappa_j \cdot Z_j$$

is used to split the node where $\kappa_j = 1$ if the variable $x_j$ does not appear as a splitting variable at any ancestor node on the path from the root to the node being split, and $\kappa_j = \kappa$ ($\kappa > 1$) if it does. This places a preference on fewer variables defining splits along such paths, and thereby defining the rules derived from the tree. In particular, once a variable $x_j$ is chosen for splitting a node, other variables that are highly correlated with it will be discouraged from splitting its descendants and thus appearing with it in the same rule. Note that this strategy does not necessarily discourage those variables that are highly correlated with $x_j$ from entering the overall predictive model $F(\mathbf{x})$ (25), (26). They are not discouraged from splitting nodes in the same tree that do not contain a split on $x_j$ at an ancestor node, and from being used for splitting in different trees. This strategy only discourages highly correlated variables from defining the same rule (not different rules) and thereby suppresses spurious interaction effects in the predictive model caused by collinearity.

The value chosen for the incentive parameter $\kappa$ should be large enough to effectively discourage spurious interactions, but not so large as to inhibit genuine interactions from entering the predictive model. It should be set to the largest value that does not degrade predictive performance as estimated by a left out test set or full cross-validation.

8.3. *Null distribution.* In order to use the statistics presented in Section 8.1 for measuring the strength of various kinds of interaction effects, one must have an idea of their value in the absence of such effects. Even if a particular interaction effect is absent from the target $F^*(\mathbf{x})$, the sample based estimate of the corresponding statistic will not necessarily be zero. Sampling fluctuations can introduce apparent interactions in the estimated model $F(\mathbf{x})$. In addition, there are types of associations among the predictor variables other than collinearity that if present can also induce spurious interactions in the model [Hooker (2004)] for which the strategy discussed in Section 8.2 is less effective.

Here we present a variant of the parametric bootstrap [Efron and Tibshirani (1993)] that can be used to derive a reference (null) distribution for any of the interaction test statistics presented in Section 8.1. The idea is to repeatedly compute these statistics on a series of artificial data sets generated from the training data, and then use the distribution of test statistic values so derived as a reference for the corresponding test statistic value obtained from the original data set.

For regression, each artificial data set is given by $\{\mathbf{x}_i, \tilde{y}_i\}_1^N$, where

$$\tilde{y}_i = F_A(\mathbf{x}_i) + (y_{p(i)} - F_A(\mathbf{x}_{p(i)})). \tag{47}$$

Here $\{p(i)\}_1^N$ represents a random permutation of the integers $\{1, 2, \ldots, N\}$ and $F_A(\mathbf{x})$ is the closest function to the target containing *no* interaction effects. For classification $y \in \{-1, 1\}$, the corresponding response values are

$$\tilde{y}_i = 2b_i - 1, \tag{48}$$

where $b_i$ is a Bernoulli random variable generated with

$$\Pr(b_i = 1) = \max(0, \min(1, (1 + F_A(\mathbf{x}_i))/2)), \tag{49}$$

and $F_A(\mathbf{x})$ is derived using (17). The "additive" model $F_A(\mathbf{x})$ can be estimated from the original training data set $\{\mathbf{x}_i, y_i\}_1^N$ by restricting the rules used in (25) and (26) to each involve only a single predictor variable. This is, in turn, accomplished by restricting the trees produced by Algorithm 1 to all have $t_m = \bar{L} = 2$ terminal nodes. Other techniques for estimating additive models could also be used [see Hastie and Tibshirani (1990)].

By construction, each data set generated from (47) or (48) has a target $F_A(\mathbf{x})$ containing no interaction effects. It has the same predictor variable joint distribution as the original training data. It also has the same (marginal) distribution of the residuals $\{y_i - F^*(\mathbf{x}_i)\}_1^N$ under the null hypothesis $F^*(\mathbf{x}) = F_A(\mathbf{x})$.

For each artificial data set $\{\mathbf{x}_i, \tilde{y}_i\}_1^N$ (47), (48), a full predictive model $\tilde{F}(\mathbf{x})$ is obtained by applying the identical procedure (modeling parameters, etc.) used to obtain the predictive model $F(\mathbf{x})$ on the original training data $\{\mathbf{x}_i, y_i\}_1^N$. The various interaction test statistics of interest obtained from $F(\mathbf{x})$ are computed on $\tilde{F}(\mathbf{x})$. The collection of these computed values over all artificially generated data sets can then be used as a reference distribution for the corresponding values obtained from $F(\mathbf{x})$, under the null hypothesis of no interaction effects in the target $F^*(\mathbf{x})$. Illustrations are provided in the data examples below.

8.4. *Discussion.* The general strategy of using partial dependence functions to detect and analyze interaction effects can be applied to any function $F(\mathbf{x})$, not just to those of the form (25) and (26). All that is required to compute partial dependence functions (40) is the value of $F(\mathbf{x})$ at various prediction points $\mathbf{x}$. Thus, this approach can be used with "black-box" prediction models derived by any method providing a way to estimate $F_A(\mathbf{x})$ (47) (49). The strategy for discouraging spurious interactions outlined in Section 8.2 can only be used with tree based methods however. Inhibiting spurious interactions can help to make the strategy more sensitive to the presence of genuine interaction effects in the target $F^*(\mathbf{x})$.

**9. Illustrations.** In this section we present applications of the interpretational tools described in Sections 6–8 to two data sets. The first is artificially generated so that results can be compared to known truth. The second is one that is often used as a test bed for evaluating prediction methods. Following Friedman and Popescu (2003), the tree ensemble generation parameters used in Algorithm 1 were $\nu = 0.01$ and $\eta = \min(N/2, 100 + 6\sqrt{N})$, where $N$ is the training sample size. The average tree size (13) was taken to be $\bar{L} = 4$ terminal nodes. Rules were derived from ensembles of 333 trees producing approximately 2000 rules used in (26) to produce the predictive model (25). These "default" parameter settings are used here for illustration; it is possible that individual results could be improved by selective tuning of some of them.

9.1. *Artificial data.* This data set consists of $N = 5000$ observations with $n = 100$ predictor variables. To somewhat realistically emulate actual data, the predictor variables were generated with ten discrete values $x_{ij} \in \{k/10\}_0^9$, with the integers $k$ randomly generated from a uniform distribution. Each response value was taken to be

$$y_i = F^*(\mathbf{x}_i) + \varepsilon_i, \tag{50}$$

where the target function is

$$\begin{aligned}F^*(\mathbf{x}) &= 9 \prod_{j=1}^{3} \exp(-3(1-x_j)^2) - 0.8\exp(-2(x_4 - x_5)) \\ &\quad + 2\sin^2(\pi \cdot x_6) - 2.5(x_7 - x_8)\end{aligned} \tag{51}$$

and $\varepsilon_i \sim N(0, \sigma^2)$ with the value of $\sigma$ was chosen to produce a two-to-one signal-to-noise ratio. Note that this target depends on only eight of the predictor variables; the other 92 are pure noise variables having no effect on the response. The coefficients multiplying each of the terms in (51) were chosen so as to give each of the first eight variables approximately equal global influence (28), (29), (35). The target function is seen from (51) to involve a strong three-variable interaction effect among $(x_1, x_2, x_3)$, a somewhat different two-variable interaction between $x_4$ and $x_5$, a highly nonlinear additive dependence on $x_6$, and linear dependencies of opposite sign on $x_7$ and $x_8$.

Applying RuleFit to these data produced a model (25) involving 351 terms (rules + linear) with nonzero coefficients. The average absolute error

$$aae = \frac{E_{\mathbf{x}y}|y - F(\mathbf{x})|}{E_{\mathbf{x}y}|y - \mathrm{median}(y)|} \tag{52}$$

was $aae = 0.49$ as estimated with 50000 independently generated test observations. The corresponding error for a model involving main effects only $(\bar{L} = 2)$ (13) was 0.61. Using only linear basis functions (24) in (25) and (26) produced $aae = 0.69$. Thus, including additive nonlinear terms in the model improves prediction accuracy by $\sim 12\%$ over a purely linear model, and allowing interaction effects produces another $\sim 20\%$ improvement. However, these prediction errors include the irreducible error caused by the additive random component $\varepsilon_i$ in (50). The corresponding errors (20) in estimating the actual target function $F^*(\mathbf{x})$ itself are respectively 0.18, 0.43, and 0.58. Thus, including interactions improved estimation accuracy by 58% over a purely additive model. Of course, with actual rather than artificially generated data, one can only estimate (52) and estimation inaccuracy on the target (20), while decreasing monotonically with (52), is unknown.

TABLE 1
*Simulated example: six most important rules—all predictions*

| Imp. | Coeff. | sup. | Rule |
|------|--------|------|------|
| 100 | 0.57 | 0.49 | $0.25 \leq x_6 < 0.75$ |
| 99 | 0.79 | 0.15 | $x_1 \geq 0.35$ and $x_2 \geq 0.45$ and $x_3 \geq 0.45$ |
| 83 | $-0.81$ | | linear: $x_7$ |
| 63 | 0.61 | | linear: $x_8$ |
| 61 | 0.34 | 0.51 | $0.35 \leq x_6 < 0.85$ |
| 58 | $-0.38$ | 0.25 | $x_4 < 0.35$ and $x_5 \geq 0.45$ |

9.1.1. *Rule importance.* Table 1 displays the six globally most important terms (28), (29) resulting from the RuleFit model (25), (26) in order of their estimated importance. Column 1 gives the respective importances scaled so that the highest estimate receives a value of 100. Column 2 shows the corresponding coefficients $(\hat{a}, \hat{b})$. For rules (7) the coefficient $(\hat{a})$ value represents the change in predicted value if the rule is satisfied ("fires"). For linear terms (24) the coefficient is its corresponding slope parameter $\hat{b}$. The third column gives the support (12), where appropriate, for the respective rules displayed in column 4.

Comparing Table 1 with the (here known) target function (51), one sees that these six most important terms (out of 351 total) provide a reasonable qualitative description of its dependence on the 100 predictor variables. None of these terms include any of the noise variables $\{x_i\}_9^{100}$. The first and fifth rules indicate larger target function values when $x_6$ is in the middle of its range of values. The second rule produces larger target values when $x_1$, $x_2$ and $x_3$ simultaneously realize high values. The third and fourth terms reflect the linear dependences on $x_7$ and $x_8$. The sixth rule indicates smaller target values when $x_4$ is small and $x_5$ is large.

9.1.2. *Input variable importance.* The upper left frame of Figure 4 shows the relative importance of the ten most important input predictor variables (35), as averaged over all predictions (28) and (29), in descending order of estimated importance. By construction, the target (51) depends on each of the first eight variables $x_1$–$x_8$ with roughly equal (global) strength and has no dependence on $x_9$–$x_{100}$. Even though the standard deviation of the irreducible error $\varepsilon$ is here one half of that of the target function, one sees that none of the 92 noise variables has estimated relative importance greater than 5% of that for the eight relevant variables.

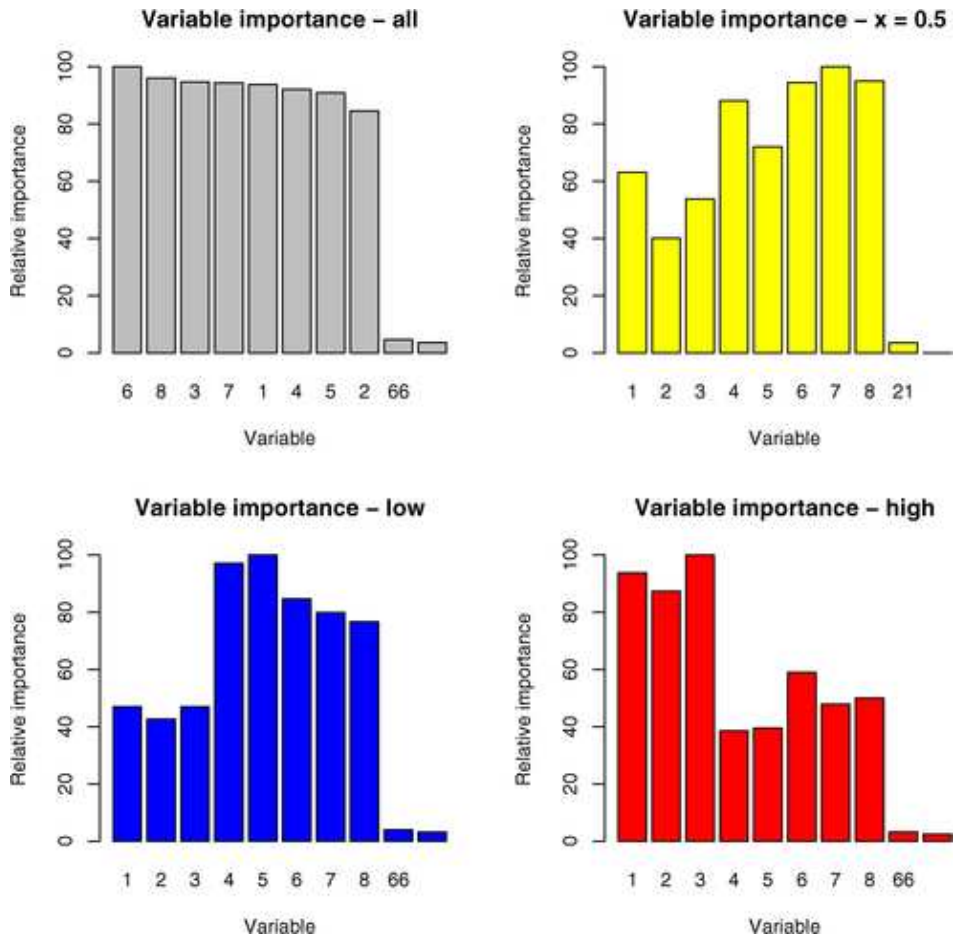

FIG. 4. *Input variable relative importances for the simulated data as averaged over all (upper left), the 10% lowest (lower left) and 10% highest (lower right) predictions, and for the single prediction point $\{x_j = 0.5\}_1^n$ (upper right).*

The upper right frame in Figure 4 shows the relative importance of the first eight predictor variables plus the two most relevant noise variables for a single prediction point $\{x_j = 0.5\}_1^{100}$ (30), (31), (35). Here one sees varying importance for each of the relevant predictor variables with the (additive) variables $\{x_6, x_7, x_8\}$ being somewhat more influential. The lower left and right frames respectively show the corresponding relative variable importances for the 10% lowest (32), (34), (35) and 10% highest (32), (33), (35) predicted target values. Here one sees that variables $x_1$, $x_2$ and $x_3$ dominately influence the highest predicted values, whereas $x_4$–$x_8$ are most influential for the lowest predictions.

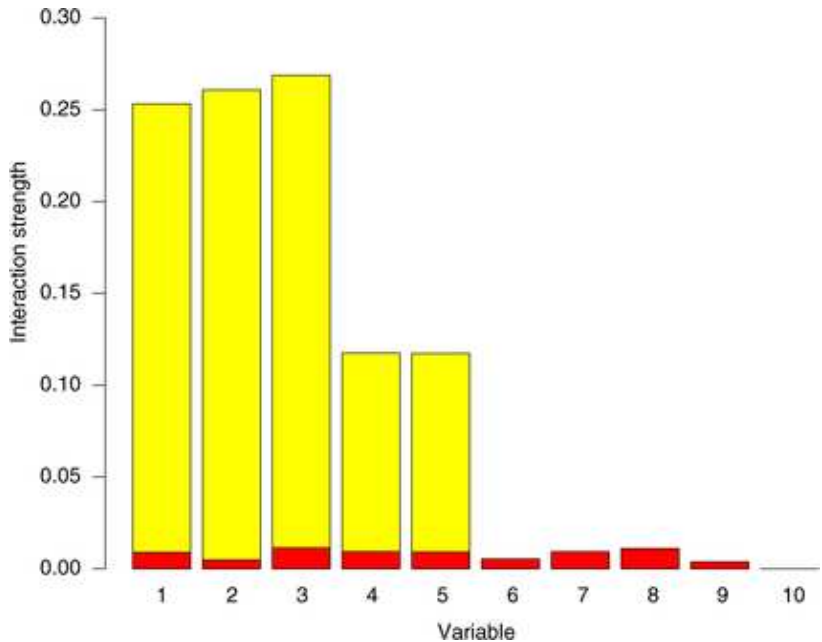

Fig. 5. *Total interaction strength in excess of expected null value of the first ten input variables for the simulated data. The lower dark bars represent the null standard deviations.*

9.1.3. *Interaction effects.* Figure 5 displays the strengths of the interaction effects involving each of the first ten predictor variables. The height of each bar represents the corresponding value of

$$\tilde{H}_j = H_j - \bar{H}_j^{(0)}, \tag{53}$$

where $H_j$ is given by (45) for each respective variable $x_j$ based on the original data, and $\bar{H}_j^{(0)}$ is the mean (null) value of the same statistic averaged over ten runs of the parametric bootstrap as described in Section 8.3. Thus, each bar reflects the value of $H_j$ in excess of its expected value under the null hypothesis of no interaction effects. The dark bars shown in Figure 5 are the values of the standard deviations $\sigma_j^{(0)}$ of the respective null distributions, so that one can visually gauge the significance of each corresponding interaction. The dark bars are plotted over the lighter ones so that the absence of a light bar indicates that the corresponding value of $H_j$ is less than or equal to one standard deviation above its null mean value $\bar{H}_j^{(0)}$.

The results shown in Figure 5 suggest that variables $x_1$, $x_2$ and $x_3$ are each heavily involved in interactions with other variables. Variables $x_4$ and $x_5$ also substantially interact with other variables, but to a somewhat lesser extent. There is no evidence of any interaction effects involving variables $x_6$–$x_{10}$.

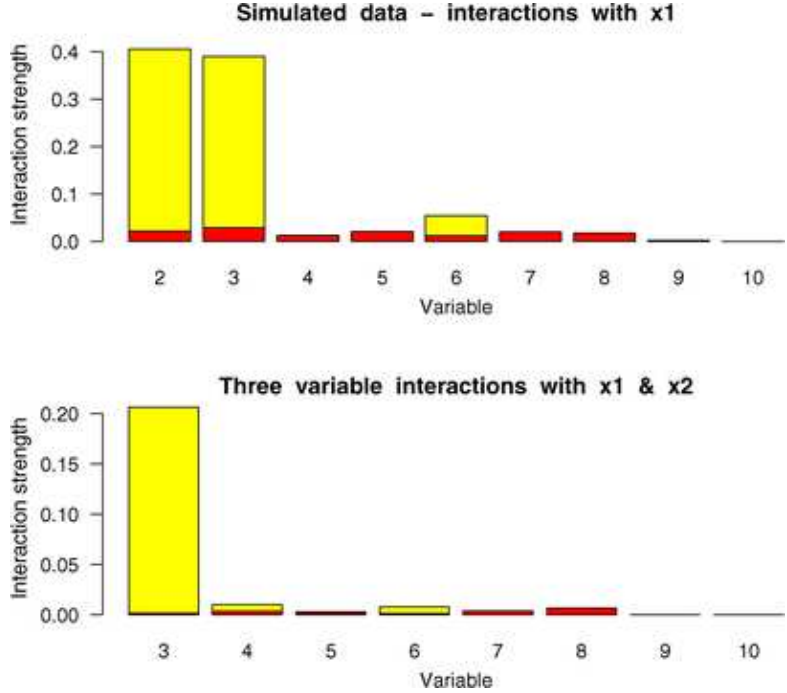

FIG. 6. *Two-variable interaction strengths of variables interacting with $x_1$ (upper) and three-variable interaction strengths of variables interacting with $x_1$ and $x_2$ (lower) in excess of their expected null value for the simulated data. The lower dark bars represent the corresponding null standard deviations.*

After identifying those variables that interact with others, it is of interest to determine the particular other variables with which each one interacts. The upper frame of Figure 6 displays the values of $\{\tilde{H}_{1k}\}_2^{10}$, where

$$(54) \qquad \tilde{H}_{jk} = H_{jk} - \bar{H}_{jk}^{(0)}.$$

Here $H_{jk}$ is given by (44) for the respective variables $(x_j, x_k)$ and $\bar{H}_{jk}^{(0)}$ is the corresponding expected null value averaged over ten replications of the parametric bootstrap (Section 8.3). The dark bars plotted over the light ones reflect the corresponding null standard deviations $\sigma_{jk}^{(0)}$.

Here one sees that $x_1$ is dominately interacting with $x_2$ and $x_3$ and there is no strong evidence of $x_1$ interacting with variables other than $x_2$ and $x_3$.

Since $x_1$ is seen to interact with more than one other variable, one can proceed to determine the orders of the corresponding interactions. The lower

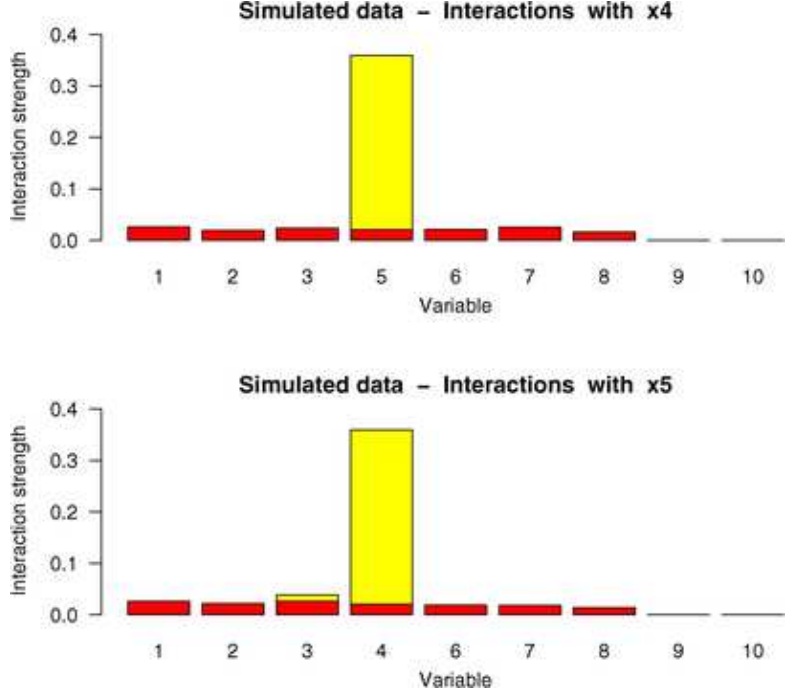

Fig. 7. *Two-variable interaction strengths of variables interacting with $x_4$ (upper) and $x_5$ (lower) in excess of expected null value for the simulated data. The lower dark bars represent the null standard deviations.*

frame of Figure 6 shows $\{\tilde{H}_{12l}\}_3^{10}$ with

$$\tilde{H}_{jkl} = H_{jkl} - \bar{H}_{jkl}^{(0)} \tag{55}$$

being the null mean adjusted analog of (46), along with $\{\sigma_{12l}^{(0)}\}_3^{10}$ (dark bars). This plot reveals that $x_1$ and $x_2$ jointly interact with $x_3$, but with no other variables, implying a three-variable interaction among these three variables but no other three-variable interactions involving $x_1$ and $x_2$.

The upper frame of Figure 7 shows $\{\tilde{H}_{4k}\}_{k\neq 4}$ (54) along with the corresponding $\sigma_{4k}^{(0)}$ (dark) for the first ten predictor variables. Here one sees that $x_4$ tends to only interact with $x_5$. The lower frame shows the corresponding interaction plot for $x_5$, which is seen to only interact with $x_4$. Thus, $x_4$ and $x_5$ interact only with each other and there is no evidence that they interact with any other variables.

The conclusion to be drawn from this analysis of interactions is that these data provide strong evidence for a three-variable interaction effect between $x_1$, $x_2$ and $x_3$, and a two-variable interaction between $x_4$ and $x_5$. There is no evidence for any other interaction effects. Note that the noise variables $x_9$ and $x_{10}$ that were judged from (35) to be irrelevant are seen to be inconsequential in the analysis of interaction effects and thus need not have been considered.

The particular target function (51) generating these data was chosen to illustrate the properties the test statistics used to uncover various types of interactions. As such, it involved strong interaction effects among some of the variables and none at all among others. Target functions occurring in practice seldom have such sharp distinctions. Often the various predictor variables tend to be involved in a wide variety of interaction effects of varying types and strength, and the goal is to uncover those (if any) that are sufficiently important.

9.1.4. *Partial dependencies.* Figure 8 displays partial dependence (40) plots on selected variables as suggested by the analysis of interactions above. For display purposes, all partial dependence functions are translated to have a minimum value of zero. The partial dependencies on $(x_1, x_3)$ and $(x_2, x_3)$ are very similar to that shown for $(x_1, x_2)$ in the upper left frame, and that for $x_8$ is very similar to that shown in the lower right frame for $x_7$ but with opposite slope. Comparing these with the actual target function (51), one sees that they provide a fairly representative pictorial description of the dependence of the response on the predictor variables.

9.2. *Boston housing data.* This is a well-known public data set often used to compare the performance of prediction methods. It consists of $N = 506$ neighborhoods in the Boston metropolitan area. For each neighborhood, 14 summary statistics were collected [Harrison and Rubinfield (1978)]. The goal is to predict the median house value (response) in the respective neighborhoods as a function of the $n = 13$ other (predictor) variables. Here we investigate the nature of the dependence of the response (measured in units of \$1000) on these predictors using the tools described in Sections 6–8.

Applying RuleFit to these data produced a model (25) involving 215 terms (rules + linear) with nonzero coefficients. The average absolute prediction error (52) was $aae = 0.33$, as estimated, by 50-fold cross-validation. The corresponding error for an additive model restricted to main effects only $(\bar{L} = 2)$ (13) was 0.37, and that for a model involving only linear terms (24) was $aae = 0.49$. Thus, the target function appears to be highly nonlinear with some evidence for interaction effects.

9.2.1. *Rule importance.* Table 2 shows the nine globally most important terms (28), (29) resulting from the RuleFit model (25), (26), in the same format as Table 1. The most important term by a substantial margin is the linear function of *LSTAT* (percent of lower status population). Its coefficient $\hat{b}$ is negative, indicating that neighborhoods with larger values of *LSTAT* tend to have lower valued homes. The linear predictor *AGE* (fraction of houses build before 1940) has a similar effect to a lesser degree.

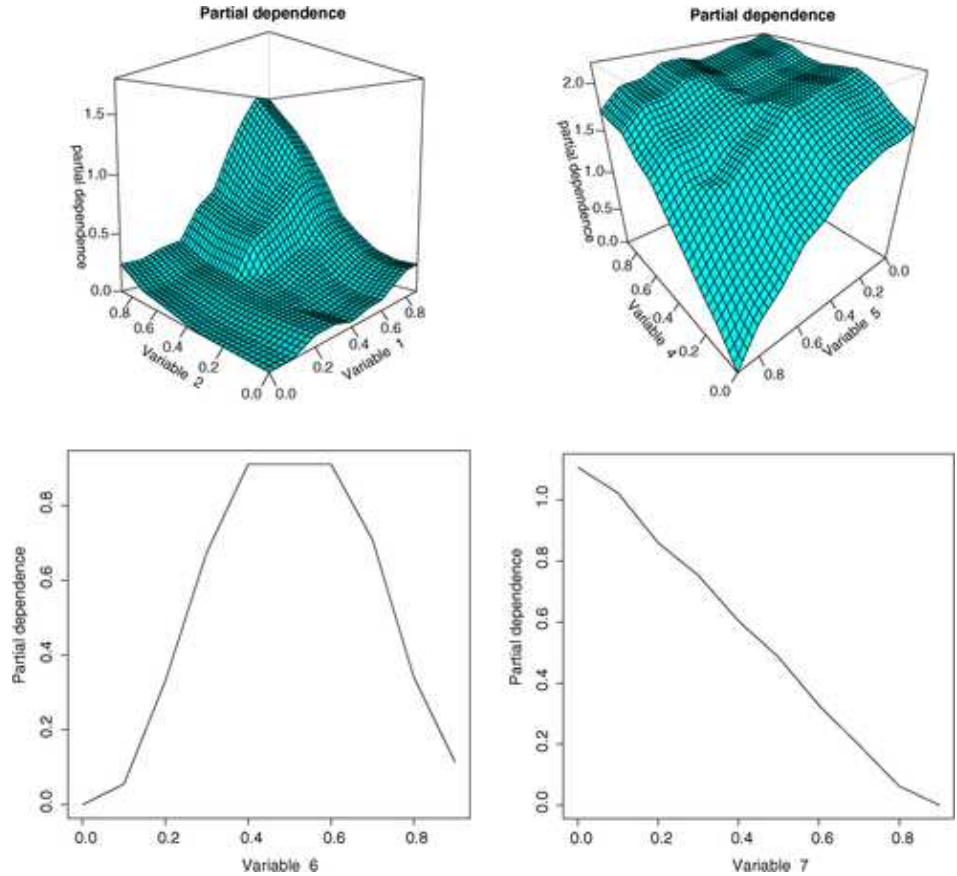

FIG. 8. *Plots of partial dependence functions on selected single variables and variable pairs for the simulated example.*

The coefficient $\hat{a}$ of the most important rule is roughly five times larger in absolute value than that of the others and indicates neighborhoods with exceptionally high housing values. These neighborhoods are characterized by being very close to Boston employment centers (*DIS*), high pupil-teacher ratio (*PTRATIO*), and very small *LSTAT*. This rule describes only five of the 506 neighborhoods: two of the six neighborhoods in Back Bay, and all three in Beacon Hill. The other rules in Table 2 indicate that neighborhoods with larger houses (number of rooms *RM*) and lower pollution (concentration of nitric oxide *NOX*), as well as larger houses and lower *PTRATIO*, tend to have higher valued homes. Neighborhoods not very close to employment centers, combined with smaller houses and higher tax rates (*TAX*), as well as combined with high *PTRATIO*, tend to have lower valued homes.

9.2.2. *Input variable importance.* The upper left frame of Figure 9 shows the global relative importances of the 13 predictor variables (28), (29), (35) averaged over all neighborhoods. In addition to those variables presented in Table 2, there is some indication that crime rate (*CRIM*) has some influence on housing values. The upper right frame shows the corresponding importances for predicting median home value in the single neighborhood comprising the town of Manchester (30), (31), (35). Here *RM* and *TAX* are relatively more influential for this prediction than on average, whereas *LSTAT* is considerably less influential. The lower left and right frames respectively show the corresponding relative variable importances for those neighborhoods with the 10% lowest (32), (34), (35) and 10% highest (32), (33), (35) predicted housing values. For the lowest predictions, the variable *LSTAT* dominates, being more than twice as important than any other variable. For the highest predicted values, *RM* is the most important variable and *PTRATIO* is nearly as important as *LSTAT*. Pollution *NOX* seems to be roughly equally relevant everywhere.

9.2.3. *Interaction effects.* Figure 10 shows the values of $\{\tilde{H}_j\}_1^{13}$ (53), along with the corresponding null standard deviations, for the Boston housing predictor variables. There is strong evidence for interactions involving *NOX*, *RM*, *DIS*, *PTRATIO* and *LSTAT*. Here we investigate further the nature of those involving *RM* and *LSTAT*.

The upper frame of Figure 11 displays the values of $\{\tilde{H}_{RM,k}\}_{k \neq RM}$ (54) along with the corresponding null standard deviations. One sees strong evidence for an interaction effect between *RM* and *NOX* and between *RM* and *PTRATIO*. The lower frame shows the corresponding plot for *LSTAT* indicating substantial interaction effects involving *LSTAT* and *NOX*, and *LSTAT* and *DIS*. Since *RM* and *LSTAT* are each seen to interact with more than one other variable, one can use (55) to investigate the presence of three-variable interactions. In this case, however, the analysis revealed no evidence for any three-variable interactions involving *RM* or *LSTAT*. This strategy can be continued to potentially uncover additional interaction effects if any.

TABLE 2
*Boston housing data: nine most important rules*

| Imp. | Coeff. | Sup. | Rule |
|---|---|---|---|
| 100 | −0.40 | | linear: *LSTAT* |
| 37 | −0.036 | | linear: *AGE* |
| 36 | 10.1 | 0.0099 | *DIS* < 1.40 and *PTRATIO* > 17.9 and *LSTAT* < 10.5 |
| 35 | 2.26 | 0.23 | *RM* > 6.62 and *NOX* < 0.67 |
| 26 | −2.27 | 0.88 | *RM* < 7.45 and *DIS* > 1.37 and *TAX* > 219.0 |
| 25 | −1.40 | 0.41 | *DIS* > 1.30 and *PTRATIO* > 19.4 |
| 20 | 2.58 | 0.049 | *RM* > 7.44 and *PTRATIO* < 17.9 |
| 19 | 1.30 | 0.21 | *RM* > 6.64 and *NOX* < 0.67 |
| 18 | 2.15 | 0.057 | *RM* > 7.45 and *PTRATIO* < 19.7 |

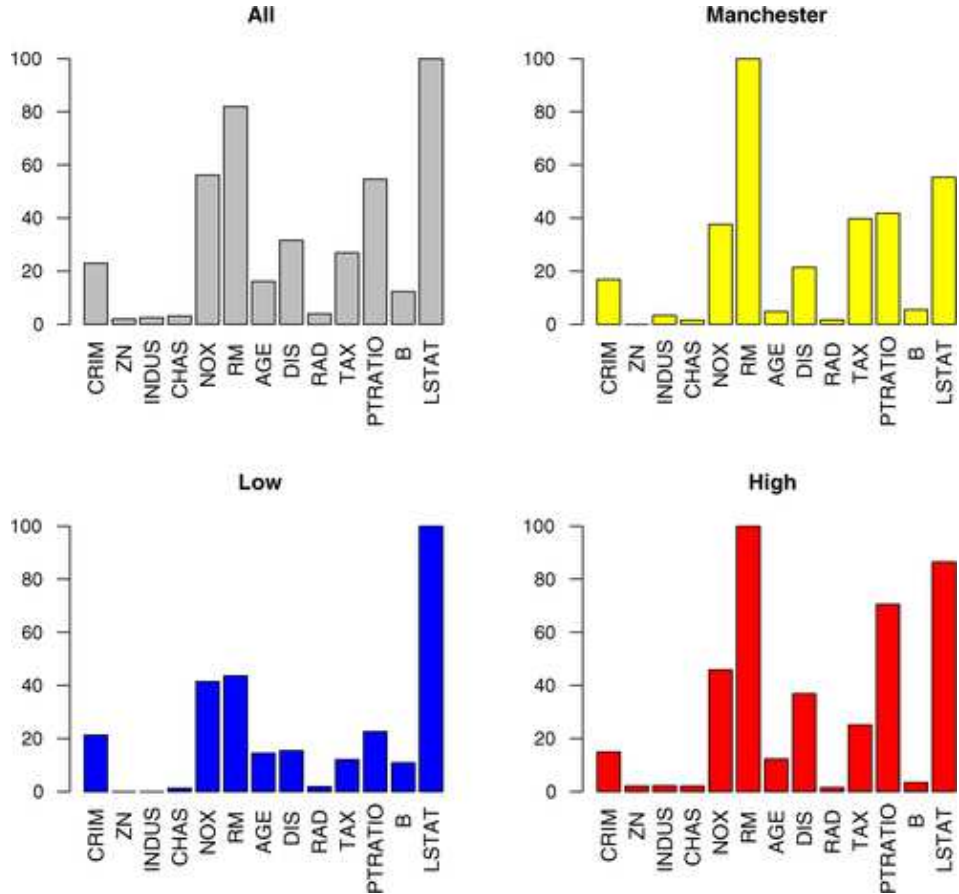

Fig. 9. *Input variable relative importances for the Boston housing data as averaged over all (upper left), the 10% lowest (lower left), and 10% highest (lower right) predictions, and for predicting the single neighborhood of Manchester (upper right).*

9.2.4. *Partial dependencies.* Figure 12 displays partial dependence functions (40) on the four variable pairs indicated above as participating in two-variable interactions. From these plots one can study the detailed nature of the corresponding interaction effects. For example, the lower right plot indicates that housing values sharply increase when LSAT and DIS simultaneously have very small values.

**10. Related work.** Predictive methods based on rules have a long history in the machine learning literature [see Mitchell (1997)]. Quinlan (1993) designed a variant of C4.5 ("C4.5 Rules") where the final model consists of a set of rules. A single large decision tree is induced and then converted to a set of rules, one for each *terminal* node. Each such rule is subsequently pruned by removing the conditions (indicator functions) that improve its estimated prediction accuracy. Finally, the pruned rules $\{r_m(\mathbf{x})\}$ are each assigned a class label and then listed in ascending order of their estimated accuracy. To obtain a prediction at a point $\mathbf{x}$, the single rule highest in this list for which $r_m(\mathbf{x}) = 1$ is used. Although there are fundamental differences, this approach is connected to the work presented here in that a decision tree induction algorithm is employed as a greedy mechanism for generating the rules.

A different rule induction paradigm used in classification context is *sequential covering*, that underlies the Inductive Logic Programming (ILP) algorithms [Lavrač and Džeroski (1994)]. The generic sequential covering algorithm induces a disjunctive set of rules by learning one rule at a time. After each rule is derived, the algorithm removes from the training data set the "positive" examples (specified $y$-value) covered by the rule. The process is then iterated on the remaining training observations. As with C4.5 Rules, the generated rule set is ordered and the single rule highest in the list that covers a point $\mathbf{x}$ is used for its prediction. Actual ILP algorithms such as CN2 [Clark and Niblett (1989)], RIPPER [Cohen (1995)], and PROGOL [Muggleton (1995)] differ with respect to the detailed techniques that implement the generic paradigm.

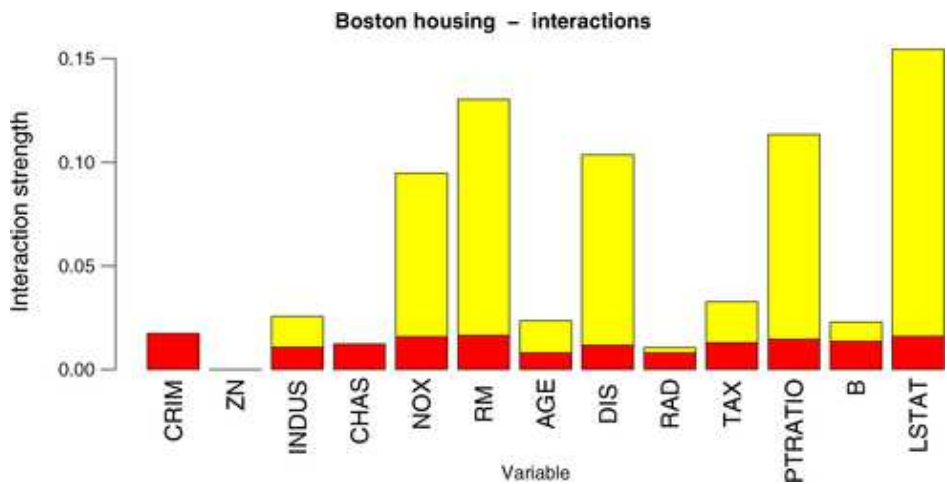

FIG. 10. *Total interaction strength in excess of expected null value of the input variables for the Boston housing data. The lower dark bars represent the null standard deviations.*

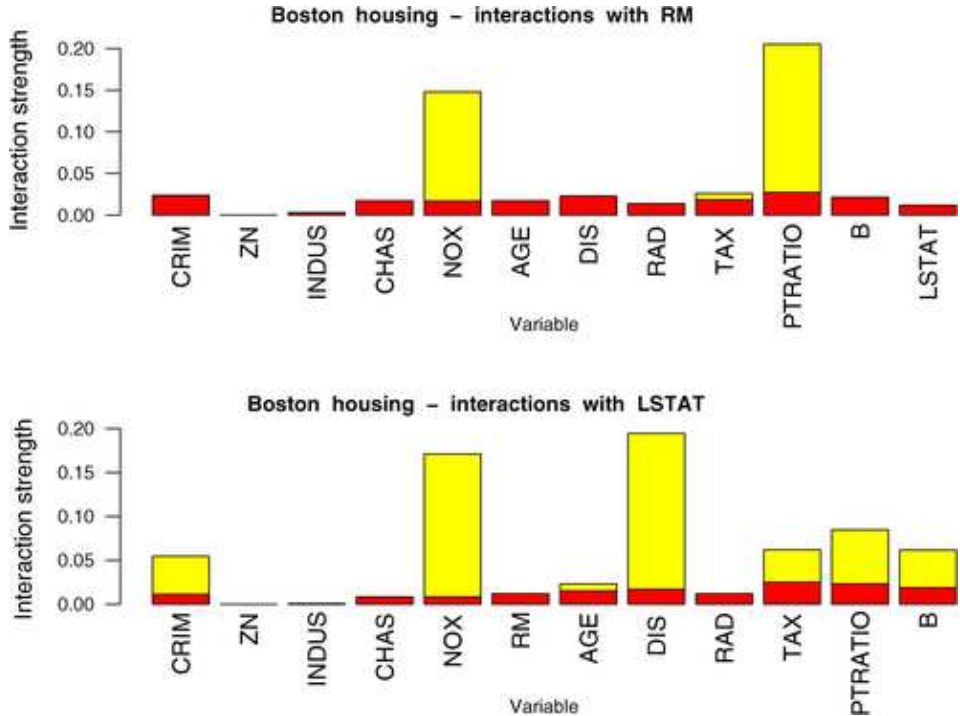

Fig. 11. *Two-variable interaction strengths of variables interacting with RM (upper) and LSTAT (lower) in excess of expected null value for the Boston housing data. The lower dark bars represent the null standard deviations.*

Although rule based, RuleFit produces fundamentally different models than the methods described above, both with respect to the methodology employed to derive the final model and the *structure* of this model. RuleFit models (25) are *additive* in rules (7) and linear terms (24) with optimized weights (coefficients), whereas the above methods produce disjunctive sets of rules using only one in the set for each prediction.

Classification ensembles that combine simple "weak" learners that are unions of conjunctive rules can be found in algorithmic implementations of the *stochastic discrimination* paradigm [Kleinberg (1996)]. Each weak learner is produced by a random mechanism (e.g., a finite union of rectangular boxes where each box is generated using a random set of variables, random centering, and random length edges). The corresponding weak learners chosen for the final model are required to satisfy certain "enrichment" and "uniformity" conditions. Details are presented in Ho and Kleinberg (1996) and Kleinberg (2000). [See also Pfahringer et al. (2004)]. As with RuleFit, stochastic discrimination combines its base (weak) learners in an additive manner. The major differences are the mechanism employed to generate the additive terms and the fact that stochastic discrimination performs a simple averaging, whereas the coefficients of RuleFit models are fit through a regularized regression (26).

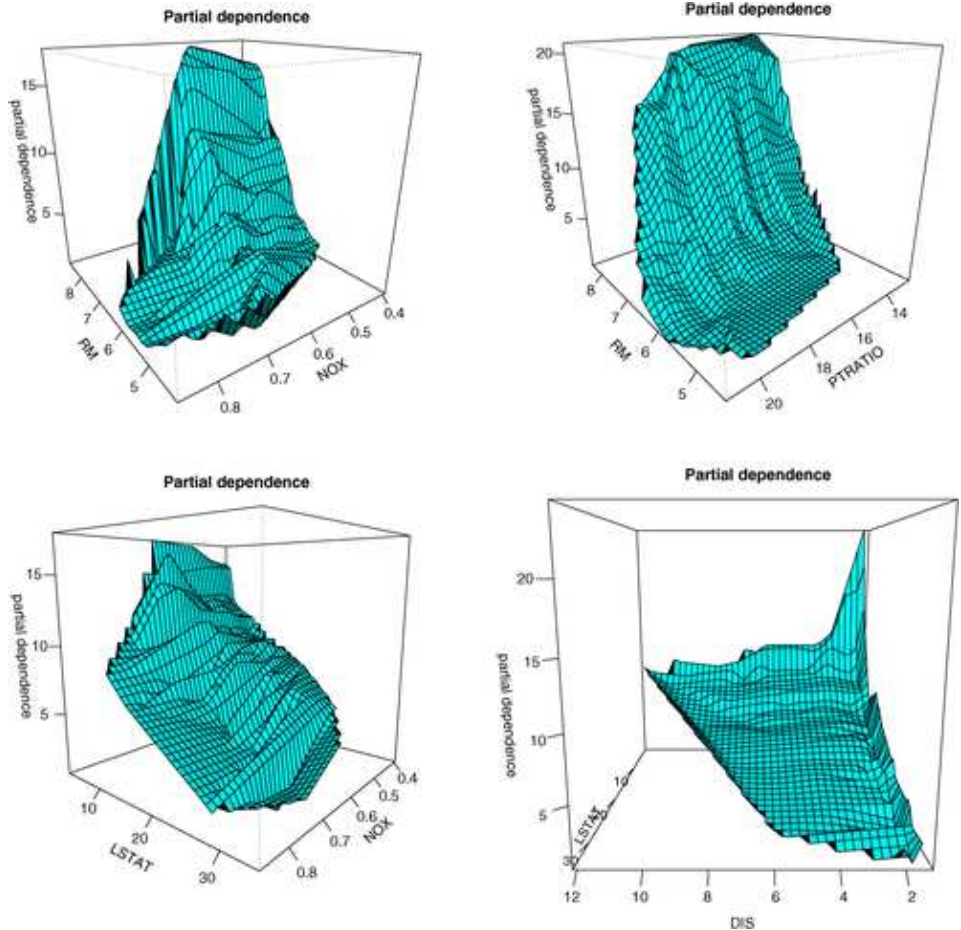

FIG. 12. *Plots of partial dependence functions on selected variable pairs for the Boston housing data.*

SLIPPER [Cohen and Singer (1999)] uses the AdaBoost strategy to produce a weighted ensemble of boosted RIPPER rules. While generally outperforming standard rule induction methods, this approach tends not to match the performance of boosted tree ensembles [Weiss and Indurkhya (2000)]. Light weight rule induction [LRI, Weiss and Indurkhya (2000)] uses a simple heuristic strategy based on the boosting concept to produce unweighted rule ensembles having an equal number of rules for each class. They provide evidence that this approach tends to outperform SLIPPER and single trees for small rule sets, and with larger ensembles was competitive with the best methods available at the time. Both SLIPPER and LRI sequentially induce relatively small rule sets, all of which are used for prediction. Rule-Fit initially induces a large number of rules and then employs regularized regression (10), (26) to produce and weight the smaller set comprising the final predictive model.

Designed for problems involving binary valued features, logic regression [Ruczinski, Kooperberg and LeBlanc (2003)] uses regression to fit the coefficients of a model that is additive in rules "logic trees." A set of admissible operations is defined for modifying the logic trees and the model building process involves randomly applying these operations in a simulated annealing context. Due to the intensive nature of the computation involved with the simulated annealing approach, logic regression can accommodate models involving relatively few logic terms. Also, generalizations to numeric and multiple-valued categorical variables complicate this approach.

Closer to the approach presented here is that of Rosset and Inger (2000). They constructed binary classification models using (unregularized) linear logistic regression where the predictors were taken to be the original input variables along with manually selected and modified C4.5 rules based on those variables.

Ruckert and Kramer (2006) propose the use of regularized regression to construct weighted rule ensembles for classification. An initial set of rules is defined without reference to the outcome variable $y$. At each step one additional rule from this set is introduced into the model in a (user) predefined order. A regularized regression is performed on the rules currently in the model, and a upper bound on the true (population) value of the fitting criterion is computed based on its empirical value and the current number of rules. These steps are repeated until all of the initial rules have been included. The model in this sequence achieving the lowest upper bound on the criterion is then used for prediction. Three fitting criteria were proposed with the *margin minus variance* (MMV), with an $l_1$ constraint on the weights being preferred among the three. Principal differences between this approach and RuleFit are that the latter uses information in the outcome variable $y$ to preferentially generate a good initial rule set (Algorithm 1), and the order of rule entry is determined by the data directly through the regularized regression procedure (10), (26).

For interpretation, Breiman et al. (1983) proposed a predictor variable importance measure for (single) trees. The relative importance for each variable was taken to be the sum of the improvements in squared-error risk on the training data at each nonterminal node split on that variable. Friedman (2001) and Breiman (2001) extended this measure to tree ensembles by simply averaging it over the trees appearing in the ensemble ("Gini" measure). Breiman (2001) also suggested a permutation based variable importance measure. The relevance of each variable was taken to be the increase in prediction risk of the model, as averaged over the training data, when the values of that variable were randomly permuted among the observations. Like those described in Sections 6 and 7, these measures reflect the marginal influence of each respective variable in the presence of the other variables. They need not reflect usefulness in the absence of other variables. Also, the permutation measure is essentially global in nature and is not readily extended to produce corresponding local measures at individual predictions (30), (31), (35). However, the Gini measure could be so extended.

Roosen (1995), Owen (2001) and Jiang and Owen (2001) study interaction effects in "black-box" models using the functional ANOVA decomposition of $F(\mathbf{x})$ and product measure. Hooker (2004) discusses the limitations of using product measure in the context of observational data and proposes alternatives that are intended to mitigate this constraint. Our approach to interactions based on partial dependence functions (Section 8.1) does not involve the functional ANOVA decomposition. Hooker (2004) observes that associations among the predictor variables can sometimes introduce distortion in partial dependence estimates based on empirical models. This motivates our approach of suppressing spurious interactions presented in Section 8.2, and using null distributions as derived in Section 8.3 to calibrate observed interaction effects.

**Acknowledgments.** We gratefully acknowledge helpful discussions with Bradley Efron and Jonathan Taylor.

## REFERENCES

BREIMAN, L. (1996). Bagging predictors. *Machine Learning* **26** 123–140.

BREIMAN, L. (2001). Random forests. *Machine Learning* **45** 5–32.

BREIMAN, L., FRIEDMAN, J. H., OLSHEN, R. and STONE, C. (1983). *Classification and Regression Trees*. Wadsworth, Belmont, CA. MR0726392

CLARK, P. and NIBLETT, R. (1989). The CN2 induction algorithm. In *Machine Learning* **3** 261–284.

COHEN, W. (1995). Fast efficient rule induction. *Machine Learning*: *Proceedings of the Twelfth International Conference* 115–123. Morgan Kaufmann, Lake Tahoe, CA.

COHEN, W. and SINGER, Y. (1999). A simple, fast and efficient rule learner. In *Proceedings of the Sixteenth National Conference on Artificial Intelligence (AAAI–99)* 335–342. AAAI Press.

DONOHO, D., JOHNSTONE, I., KERKYACHARIAN, G. and PICARD, D. (1995). Wavelet shrinkage: asymptotia? (with discussion). *J. Roy. Statist. Soc. Ser. B* **57** 301–337. MR1323344

EFRON, B. and TIBSHIRANI, R. (1993). *An Introduction to the Bootstrap*. Chapman and Hall, New York. MR1270903

FREUND, Y. and SCHAPIRE, R. E. (1996). Experiments with a new boosting algorithm. *Machine Learning: Proceedings of the Thirteenth International Conference* 148–156. Morgan Kauffman, San Francisco.

Friedman, J. H. (2001). Greedy function approximation: A gradient boosting machine. *Ann. Statist.* **29** 1189–1232. MR1873328

Friedman, J. H. and Hall, P. (2007). On bagging and nonlinear estimation. *J. Statist. Plann. Inference* **137** 669–683.

Friedman, J. H. and Popescu, B. E. (2003). Importance sampled learning ensembles. Technical report, Dept. Statistics, Stanford Univ.

Friedman, J. H. and Popescu, B. E. (2004). Gradient directed regularization for linear regression and classification. Technical report, Dep. Statist. Dept. Statistics, Stanford Univ.

Harrison, D. and Rubinfield, D. C. (1978). Hedonic prices and the demand for clean air. *J. Environmental Economics and Management* **8** 276–290.

Hastie, T. and Tibshirani, R. (1990). *Generalized Additive Models*. Chapman and Hall, London. MR1082147

Hastie, T., Tibshirani, R. and Friedman, J. H. (2001). *Elements of Statistical Learning*. Springer, New York. MR1851606

Ho, T. K. and Kleinberg, E. M. (1996). Building projectable classifiers of arbitrary complexity. In *Proceedings of the 13th International Conference on Pattern Recognition* 880–885. Vienna, Austria.

Hooker, G. (2004). Black box diagnostics and the problem of extrapolation: extending the functional ANOVA. Technical report, Dept. Statistics, Stanford Univ.

Huber, P. (1964). Robust estimation of a location parameter. *Ann. Math. Statist.* **53** 73–101. MR0161415

Jiang, T. and Owen, A. B. (2001). Quasi-regression for visualization and interpretation of black box functions. Technical report, Dept. Statistics, Stanford Univ.

Kleinberg, E. M. (1996). An overtraining-resistant stochastic modelling method for pattern recognition. *Ann. Statist.* **24** 2319–2349. MR1425956

Kleinberg, E. M. (2000). On the algorithmic implementation of stochastic discrimination. *IEEE Trans. Anal. Machine Intelligence* **22** 473–490.

Lavrač, N. and Džeroski, S. (1994). *Inductive Logic Programming: Techniques and Applications*. Ellis Horwood.

Mitchell, T. (1997). *Machine Learning*. McGraw-Hill, New York.

Muggleton, S. (1995). Inverse entailment and PROGOL. *New Generation Computing* **13** 245–286.

Owen, A. B. (2001). The dimension distribution and quadrature test functions. *Statist. Sinica* **13** 1–17. MR1963917

Pfahringer, B., Holmes, G. and Weng, C. (2004). Millions of random rules. In *Proceedings of the 15th European Conference on Machine Learning (ECML/PKDD 2004)*. Morgan Kaufmann, San Mateo.

Quinlan, R. (1993). *C4.5: Programs for Machine Learning*. Morgan Kaufmann, San Mateo.

Roosen, C. (1995). Visualization and exploration of high–dimensional functions using the functional Anova decomposition. PH.D. thesis, Dept. Statistics, Stanford Univ.

Rosset, S. and Inger, I. (2000). KDD–CUP 99: Knowledge discovery in a charitable organization's donor data base. *SIGKDD Explorations* **1** 85–90.

Ruckert, U. and Kramer, S. (2006). A statistical approach to learning. In *Proceedings of the 23rd International Conference on Machine Learning.* Morgan Kaufmann, San Mateo.

Ruczinski, I., Kooperberg, C. and LeBlanc, M. L. (2003). Logic regression. *J. Comput. Graph. Statist.* **12** 475–511. MR2002632

TIBSHIRANI, R. (1996). Regression shrinkage and selection via the lasso. *J. Roy. Statist. Soc. Ser. B* **58** 267–288. MR1379242

WEISS, S. and INDURKHYA, N. (2000). Lightweight rule induction. In *Proceedings of the 17th International Conference on Machine Learning* (P. Langley, ed.) 1135–1142. Morgan Kaufmann, San Mateo.

DEPARTMENT OF STATISTICS
STANFORD UNIVERSITY
STANFORD, CALIFORNIA 94305
USA
E-MAIL: jhf@stanford.edu
    bogdan@stat.stanford.edu